# Ernest Elliott Markwick: variable stars and military campaigns

Jeremy Shears

**Abstract**

Colonel E.E. Markwick, CB, CBE, FRAS (1853 – 1925) pursued a distinguished career in the British Army, serving in Great Britain and other parts of the Empire and rising to the rank of Colonel. He was an original member of the BAA and went on to become President between 1912 and 1914. His main observational interest was the study of variable stars and he independently discovered two variables, RY Sgr and T Cen. He directed the BAA Variable Star Section from 1899 to 1909, organising its work along lines that are largely pursued even to this day and which other variable star organisations around the world have emulated.

**Introduction**

Ernest Elliott Markwick (1853 – 1925; Figure 1) was a founding member of the BAA and served as President from 1912 to 1914. His main astronomical interest was the observation of variable stars, although he was also an active solar observer. The British Astronomical Association's Variable Star Section (BAA VSS), launched in 1890, is the world's longest established organisation for the systematic observation of variable stars (1). Markwick became its second Director in 1899 and remained in office until 1909. He transformed the work of the VSS and organised it along the lines that would be familiar with today's VSS members. Markwick led a distinguished Army career, serving in southern Africa, Gibraltar, Ireland and England, ultimately attaining the rank of Colonel.

In this paper I describe Markwick's life and his contributions to astronomy, with a particular focus on variable stars. I have drawn on Markwick's own writings, for he was a prolific author both of informal notes and letters, such as those appearing in the *English Mechanic* (2), as well as formal scientific papers. I also referred extensively to three obituaries; for the BAA (3), the RAS (4) and for readers of the *English Mechanic* (5).

**Early life and Army career**

Markwick was the eldest son of William and Eliza Markwick (6) and was born in Kent on 19 July 1853. He attended King's College School, London, from March 1870 to December 1871 (7) where he excelled academically, being awarded the Class and the Mathematics Prizes in the Upper Fifth Form (8). In 1872 he began his Army career by joining the Control Department which later became the Army Ordinance Department and which was responsible for the supply and maintenance of weaponry, munitions and other equipment. His first overseas posting was to southern Africa where he saw much action including during the Zulu War of 1879, the First Anglo-Boer War of 1880-81, when he was mentioned in despatches during the siege of Pretoria, and the Bechuanaland Expedition of 1884-85 (9). The Bechuanaland Expedition comprised four thousand troops that were led by Major General Charles Warren (1840-1927) north from Cape Town to exert British sovereignty over Bechuanaland (modern day Botswana) in the face of encroachments from Germany, the Transvaal and various Boer states. The expedition achieved its aims without bloodshed (10). Markwick was again mentioned in despatches and received a special promotion (9) .

As with many people taking a sea voyage to the southern hemisphere, Markwick was struck by the changing aspect of the night sky as he sailed from Great Britain towards southern Africa with well-known constellations slipping below the northern horizon and unfamiliar ones coming into view in the south (11):

"I shall not forget the night on the voyage out from England when the clouds of Magellan were first seen, an earnest of what was to come when still lower latitudes were reached."

Once he reached his destination he discovered that the night skies were often highly conducive to star gazing and remarkable for their transparency. For example, whilst in the Transvaal in 1881 he "on one night noted Mercury and Venus set, both of them disappearing sharply at the horizon" (12). In later writings, he frequently referred to the dark clear skies that he enjoyed especially on his travels though the dry grasslands of the *highveld* (Figure 2):

"In the interior and uplands of South Africa the winter season was nearly always dry and bright. Every day a brilliant sun, and at night a perfectly clear sky, with cold, frosty air. This state of things continued for months at a time without any change..... The three features of the nightly sky which I think strike the visitor to southern climes most are the Milky Way, Zodiacal Light, and Clouds of Magellan."

Although he carried a small telescope with him, many of his observations were made with the naked eye:

"Many is the time when on the march, or "on trek ",.........I have rolled myself up at nightfall in blanket and rugs, for one bivouacks [*sic*] with ease in the winter owing to the dryness of the atmosphere, and taken a last peep at the lovely stars overhead, to wake next morning before dawn and see a wondrous change in the position of the heavenly bodies. Pages could be penned about these ever-glorious skies"

The Zodiacal Light was a particular favourite: "it is a wondrously beautiful spectacle, seen night after night at a certain period of the year. It got to be quite a common sight, and I always found that the axis coincided approximately with the Ecliptic."

But a soldier's life in southern Africa was not without hazards. In addition to the dangers of the various military campaigns, with the ever present threat of attack from massed Zulus or Boer guerrillas, there were natural hazards to contend with too. A particular concern during the rainy season in Natal was thunderstorms:

"The lightning was very dangerous, and there were many cases of accident from it....... On one occasion a tent at Fort Napier, Pietermaritzburg, was struck, and I saw the soldier's waist belt and bayonet which had been hanging from the pole scarred and damaged." (13)

Markwick returned to England in 1885 and his next major postings were as Chief Ordnance Officer in Cork (1887-93), Gibraltar (1893-1898), Western District (1898-1903), based in Devonport, and the $2^{nd}$ Army Corps (1903-04). In 1905 he became Assistant Director of Ordinance Stores, Southern Command, retiring the same year and being made Companion of the Order of the Bath (Military).

Whilst in Gibraltar he gave a talk (14) about the night sky to "an audience largely consisting of soldiers" (i.e. the lower ranks rather than Officers) during which he "pointed out that our sentries at night..........might often make observations which would be of interest to themselves and others". Following the talk he was brought to book by a high ranking officer who, only partly in jest, said that the soldiers should be "keeping watch on terrestrial matters". Reflecting the social *mores* of the Victorian era and a highly class-conscious society, Markwick went on to say "I fear, however, that there is but little chance of their [i.e. the soldiers] indulging in such a recreation, for astronomy is to that class, usually 'caviare to the general'" (15). He went on to redeem himself, at least to the eyes of the modern reader, by saying "there are, however, exceptions, and I have sometimes come across men who took an interest in the subject, and were delighted to have a peep through an astronomical telescope".

In 1905, having lived in a variety of places during his Army career, Markwick retired to Boscombe (16) on the south coast of England. However, his seaside retirement, during which he planned to devote himself to his many interests, was not to last. At the outbreak of the First World War in the summer of 1914, he volunteered to serve King and country and was appointed Assistant Director of Ordnance Stores at Headquarters, Irish Command in Dublin. He remained in that capacity for the duration of the war and was finally demobilised in 1919. On 1 January 1919 he was honoured as Commander of the Military Division of the Most Excellent Order of the British Empire (CBE) in recognition of "valuable services in connection with the War" (17). His second retirement was spent in West Moors, Dorset (18).

Markwick married in 1882, Amy (19), the only child of F.M. Murton of Natal and they had two sons and a daughter.

**A developing interest in astronomy**

So when did Markwick's interest in astronomy begin? His earliest astronomical recollection was of Donati's comet (C/1858 L1), one of the most impressive comets of the $19^{th}$ century, which he saw "from Plymouth Hoe on a starry night in 1858" at the tender age of 5 (20). In his late teens he read a series of popular articles on various astronomical topics written by Sir John Herschel and he dates his "conversion to the charms of Urania from about that time" (21). As his interest in astronomy developed he read constantly, including the *Ipswich Lectures* (22) given by G.B. Airy (1802-1892), Norman Lockyer's (1836-1920) *Elementary Lessons* and various books by the astronomy populariser R.A. Proctor (1837-1888) (23).

His first telescope was a refractor "by Dolland with an object glass from 1½ to 2 inches [3.8 to 5 cm] in diameter, of good defining power", giving an erect image. The instrument was unmounted and its weight was not conducive to astronomical observations. On his 21$^{st}$ birthday his parents presented him with a 2¾ inch [7 cm] refractor and his astronomical observations began in earnest on that very day, 19 July 1874, "working up from the astonished star-gazer's casual peeps to perhaps something like a little systematic work" (20). For many years this was his only telescope and, even when he obtained other instruments, he continued to use it pretty much for the rest of his life (24). Its size and portability meant that it could go with him on his various Army postings and expeditions. He had a great affection for the small telescope describing it, in 1892, almost as a friend (20):

"This telescope has made a voyage to the Southern hemisphere and back again. With it have been observed, *inter alia*, a transit of Venus, a transit of Mercury, and the "Great Comet" of 1882 [C/1882 R1]..........It has been transported on the shoulders of lusty Kaffirs, and gone through perils by land and perils by water, so much so that it has the appearance, externally, of a very old veteran. To my certain knowledge it has twice been knocked down when standing, and has been in hospital on several occasions. In fact the calculation of error due to "flexure" in this instrument would present a very elegant problem to a computer who was *not* paid by piece-work. Yet to all intents and purposes the old glass yet remains as good as ever, the definition always sharp..."

The telescope came equipped with that dreaded device that accompanied many a small Victorian telescope: the pillar and claw stand (25), which Markwick soon found to be completely unsatisfactory for astronomy. He replaced it with a home-made altazimuth tripod, which in turn was superseded by "a roughly equatorial stand of iron, mounted on wooden legs" and subsequently by a Wray equatorial and tripod. The instrument can be seen in Figure 3, whilst Figure 4 shows the small roll-off roof observatory which he had built for it whilst based in Devonport around the turn of the century. Permission was granted by the War Office to erect the structure, providing that he would remove it should they request it at any time (26)

A little while after the trusty 2¾ inch refractor came into his possession, Markwick obtained a pair of 1½ inch (38 mm) binoculars having a magnification of 5. Although he noted that such "an instrument....would perhaps been looked upon with contempt in earlier days" it actually yielded many of his variable star observations over the years.

In the 5 years following the arrival of the 2¾ inch refractor, Markwick tracked down most of the double stars and nebulae listed in T.W Webb's (1807-1885) *Celestial Objects for Common Telescopes* (27). Whilst he enjoyed this form of celestial tourism and he considered it to provide a good training ground in astronomical observation, he was aware that his observational work ultimately lacked focus. Like so many amateur astronomers starting out in the hobby, he "used to potter about among the stars, flitting from one pair to another, from one nebula to another, under the guidance of dear old Webb" (28). With hindsight he wished he had perhaps been a little more purposeful: (20)

"One looks back with pleasure to the many enjoyable hours spent in exploring the beauties of the sky under the guidance of the ever interesting and conscientious Webb. And perhaps this was not without its value in gradually bringing one into more habits of observation..........But the general impression remains, after a perusal of these old notes, that the work should have been more condensed. More careful estimates of magnitudes and positions might have been made. The hour of observations was constantly omitted....and often the more experienced eye detects a vagueness in description which should not be there".

In later years, after he retired to Boscombe (29), he owned an 8½ inch (21.6 cm) altazimuth Calver reflector and a 4 inch equatorial Grubb refractor.

**The RAS and Presidency of the BAA**

At about the same time as he completed his survey of Webb's objects, Markwick applied for Fellowship of the Royal Astronomical Society, which was granted on 14 March 1879 (Figure 5). His application, was supported by Arthur Cowper Ranyard (1845-1894; Figure 6), RAS Secretary, "from personal knowledge". Markwick had developed a friendship with Cowper Ranyard when he sought the latter's advice on solar observing (30). Other notable astronomers also signed his application form (31): Andrew Ainslie Common (1841-1903; Figure 7) and Herbert Sadler (32) (1856-1898).

Markwick was also a member of the Liverpool Astronomical Society (LAS), which had developed into a national organisation, often regarded as the forerunner of the BAA. He was elected to the LAS Council in 1887 (33), serving under the President, W.F. Denning (1848-1931). When it was proposed by W.H.S Monck (1839-1915) in a letter to the *English Mechanic* on 18 July 1890 that a national amateur astronomical association with headquarters in London be set up (34), Markwick offered his wholehearted support for what soon became the BAA, writing: (35)

"I for one shall be very pleased to join such an association as has been sketched by Mr. Monck in his very excellent letter on the subject. It is evident that there is already a consensus of opinion in favour of starting it, and I am confident that there is a very wide field of action open to it. The sooner it is begun the better".

Markwick pointed out that the LAS was no longer serving its members properly, especially those located far from Liverpool, as it had ceased to publish a journal, a situation which he hoped the new society would rectify. He commented that the papers in the Monthly Notices of the RAS were "above the work of most amateurs, and there is ample room for the existence of the proposed association" (35).

When E.W. Maunder (1851-1928) formally announced the setting up of the new British Astronomical Society (later changed to Association) in August 1890, Markwick was listed as a member of the provisional committee (36). The first general meeting was held in London on 24 October 1890, when the Officers and Council were elected. Markwick was not amongst them, although he was an original member of the Association, presumably due to the fact that he was based in Ireland and rarely travelled to London.

Markwick's first JBAA paper was on "Naked Eye Astronomy" published in the fourth edition of the *Journal* (37). This was a clearly written guide for members on how to get started in astronomy and was aligned with his desire that the BAA should cater not only for those people who wanted to do regular observational work, but also those who had more of a passing interest in the science, perhaps more as a recreational activity. He noted that a "lady who has recently joined our Association remarked to [me], with reference to the first number of the Journal, which had just appeared, that its contents were somewhat too abstruse for her; she would like something simpler and not so advanced" (38).

Markwick went on to serve as BAA President from 1912 to 1914. He was invited to take the Chair for the first time at the BAA meeting held in Sion House, London, on 30 October 1912, by the retiring President, Edward B. Knobel (1841-1930; Figure 8). Markwick told the members present at the meeting that he aimed to be guided by two mottos: "Honour to Astronomy" and "Loyalty to the Association" (39). He gave his first Presidential Address on 29 October 1913 (30). His subject was "The Mission of the British Astronomical Association" and he reviewed the observational programmes, and their successes, of the BAA's observing sections.

By the time of Markwick's second Presidential Address on 28 October 1914, the world had changed and a calamity of unprecedented proportions was evolving in the form of the First World War. As noted previously, Markwick had volunteered for active service during the summer and had been posted to Ireland, as a result of which he was unable to attend in person. The Association's Vice President, the Astronomer Royal, F.W. Dyson (1868-1939: Figure 9), read a letter from Markwick which began (40):

"Ladies and Gentlemen, owing to my having been called up at the recent mobilisation to enter the duties of a post on the Staff of a large Command, I have been able to devote little, or rather no, time recently to astronomical matters.

Consequently I am not in a position to make such remarks as I should have liked on the history of the Association during the past Session".

Dyson went on to read Markwick's Presidential address on "Stellar Variation" which Markwick had "prepared some months ago", commenting that he was "very glad to be of the slightest service to one who was serving his King and country in that anxious time" (40). Although the popular view in the country in the early months of the war was that it "would be over by Christmas", this was not to be and, because of his military commitments, Markwick did not attend another BAA meeting for five years, the next occasion being the AGM of 30 October 1919 (41).

**Variable star observations and discoveries**

Quite when Markwick began observing variable stars in earnest is not known. His first published variable star estimate was of η Car in 1883 which he made whilst in southern Africa (42) (43). The main hurdle he, and other variable star observers of the time, encountered in their work was the lack of reliable comparison star sequences. In many cases, he had to make his own sequences, using magnitudes given in a variety of star atlases and catalogues. The situation began to change for the better when Edward C. Pickering (1846-1919; Figure 10), Director of Harvard College Observatory (HCO), published the first Harvard Photometry in 1884 and the two exchanged letters over the subsequent years (3) (44).

The BAA Variable Star Section database lists 7774 observations by Markwick (45), the first one being of U Mon on Christmas day 1888. This is certainly an underestimate as not all his observations have been entered into the database, including those pre-dating 1888 and some southern variables that were not on the VSS programme list (46). An example of Markwick's log book entries showing observations of the Algol-type eclipsing variable, u Her, is shown in Figure 11.

Whilst in southern Africa in the early 1880s, he set about a systematic survey of the southern skies. His aim was to compare stars which were plotted on his copy of Gould's *Uranometria Argentina* (UA) (47) with the sky with the aim of identifying new variable stars (48):

"Armed, then, with this work, we proceeded with youthful zeal to explore the region from 75⁰ S. Dec. to the S. Pole. Every star was examined with a small hand telescope of 1¾ in. [4.4 cm] aperture, and an erecting eyepiece. The result was that only a very few discrepancies in brightness, and those of a trifling nature, came to light, to which, may be added one or two slight errors in the atlas itself. Here the search ended for a time, to be changed for the equally fascinating pursuit of searching for comets, none of which, unfortunately, fell into our net".

On moving back to the UK "a second search was started in 1885, embracing the zone from 15° S. Dec to 10° N. Dec. Every star in that zone contained in the U.A., right round the heavens was examined with a binocular in the years 1886-1890. No new variable was discovered, but certain discrepancies in magnitudes were recorded" (48).

His Gibraltar posting, being located at 36⁰ N, allowed him to reacquaint himself with some of the southern skies that he had enjoyed in southern Africa (49): "So far as my experience has gone, one gets some very clear skies here, and one can work with the binocular among 7 m[ag] stars down to Dec. -45° with ease. I was pleased the other night to recognise an old friend in the shape of Omega Centauri". During his systematic search he identified many suspect variables including independently discovering two important variable stars: RY Sgr and T Cen.

RY Sgr is a member of the small group of R CrB stars. The light curves of R CrB stars show dramatic and unpredictable fades of several magnitudes that occur within a few weeks. Over succeeding months, they gradually recover their original brightness. The surfaces of R CrB stars are unusually poor in hydrogen, but rich in carbon and nitrogen, which implies that they are highly evolved. The fades are now known to be caused by condensation of carbon, making the star fade in visible light, while measurements in infrared light exhibit no real luminosity decrease. Markwick noted RY Sgr as 7$^{th}$ magnitude in July 1893, but not seen in his binoculars (thus fainter than mag 9) in September (50). E.C. Pickering described the events surrounding the discovery, including how it nearly "got away", which was independently made by Markwick and Williamina Fleming (51) (1857-1911) at HCO in the HCO Circular no. 7 dated 5 June 1896 (52):

"The star -33⁰ 14076 [now known as RY Sgr (53)] is a very remarkable object. It was one of a list of 42 stars suspected of variability, sent here for examination, by Col. Markwick. A report was sent to him that an examination of several photographs failed to show any sign of variability. A few days later an object having a peculiar spectrum was discovered by Mrs. Fleming. All the plates of the region were examined and its variability established. It was about to be published in [HCO] Circular No. 6, when it was found to be identical with the star of Colonel Markwick. It was accordingly reported to him for announcement, but he has kindly authorised its publication here".

It was with some degree of satisfaction that Markwick, quite rightly, realised that his modest search had yielded a result comparable to that achieved by the well-resourced HCO with its photographic patrol and supporting team of researchers (48):

"[I] confess at the time to a mild feeling of exultation that our small "regulation pattern" binocular (price two guineas) was partly the means of bringing this variable to light"

Markwick's chart of the area around RY Sgr is shown in Figure 12 and Figure 13 shows the light curve of the variable between 1996 and 2002, showing a deep fade from a mean of ~6.4 mag to 11.5 mag.

By contrast, Markwick's other discovery T Cen is a Mira-type long period variable which has one of the shortest known periods of this type. He described the discovery circumstances thus: (54)

"When comparing...the *Uranometria Argentina* with the sky on 1894, May 6, and using a binocular magnifying about 5 times, I noticed the star No. 252 of Centaurus (rated in that work at 6.9 mag.) looked about 8.4 or 8.5 mag....On May 26 I was surprised to see it at...6.25 [mag]".

He first announced his discovery in the Gibraltar Chronicle of 13 July 1894 (55) and then in the *English Mechanic* of 3 August 1894 (56). A more detailed description of his discovery was given in a JBAA paper in early 1895 (54). Given the rather limited circulation of these organs outside their respective countries of origin, it was perhaps not surprising that Pickering announced the independent discovery of the same star on 12 December 1895 (57). Having read Pickering's paper, Markwick set out his claim for priority in a letter to *Astronomische Nachrichten* (55). Plots of Markwick's observation of T Cen from the 1894 and 1895 observing seasons are shown in Figure 14. The General Catalogue of Variable Stars lists the range of T Cen as mag 5.5 to 9.0 with a mean period of 90.44 days (58).

Markwick made two further independent discoveries of long period variables during the course of his Gibraltar survey (4): S Scl, with a range of mag. 5.5 to 13.6 and a period of about a year, and RR Sgr, with a range of 5.4 to 14.0 and a period of about 336 days (58). In the case of S Scl, he gracefully conceded that this time he was indeed pipped at the post by the HCO team (59):

"I commenced these observations [of S Scl] in entire ignorance that anyone else was watching the star, until my attention was drawn to *Astronomische Nachrichten* No. 3225 p.161 [reference (60)] in which the announcement was made of a new variable star close to the above mentioned position".

**Director of the Variable Star Section**

The BAA Variable Star Section was launched at the end of 1890 with John Ellard Gore (1845-1910), an amateur astronomer whose research on variable and double stars was highly regarded (61), as its first Director. By the end of 1891 he had recruited 11 members, including Markwick (1). The main work of the Section was directed toward the discovery of new variable stars and novae, with the observation of known variables being a smaller part (62) (63). Several Section reports were published, but they were mainly lists of separate observations made by the observers, rather than analyses of the observations. Unfortunately Gore did not attempt to encourage cooperation between members on specific objects – observers decided which stars they wanted to observe. This was in fact the way most variable star observers worldwide operated at that time, independently accumulating, analysing and publishing their own observations (64). During the 1890s the VSS began to dwindle both in enthusiasm and number of observers. Whilst Gore was not the most dynamic of leaders, he was nevertheless working within the expectations of a section Director as laid down by the BAA Council. Gore retired as VSS Director in 1899 and Markwick was the obvious successor. He received a formal request from the BAA Council to take over the position and his consent was reported to BAA members assembled at the Association's monthly meeting of 27 December 1899 which was received "to much applause" (65). Markwick commented in his typical self-deprecating style: (66)

"I have agreed, although with a certain amount of diffidence, partly on account of becoming successor to one who is a past master both in the observation and literature of variable stars (and to whom the writer owes a debt of gratitude for much help, encouragement, and advice in this very line of work, in past years), and partly on account of being called on to "direct" some who may probably know a great deal more on the subject than the Director. A third reason which weighed against acceptation of the post was the fact of having only spare hours in which to attend to the affairs of the Section..............the first two objections are my misfortune; the third I hope to meet by strict "attention to business", even if only in spare time".

Markwick was certainly an inspired and popular choice. Not only was he a well-known variable star observer, but he soon brought his personal drive, enthusiasm and organisational skills, honed through years of military service, to bear on Section matters. One of the first strategic changes he made was to concentrate the efforts of observers on a limited programme of stars. From his own practical experience, he realised it was far better for the Section to obtain good observational coverage of a few stars, noting that "[e]xperience shows that that the amateur makes his working list far too large" (66). Thus the first list contained just 12 stars (Table 1), although the list was expanded to 46 at the end of 1900 and even further in later years. His military experience also taught him that all "soldiers" needed a call to arms, which he duly issued (66):

"Now is the time to rally round our standard and start a series of observations which will bear fruit in the future in the shape of a great increase to our knowledge of the variables, for the work of observing these is essentially such as lends itself to cooperation".

The theme of cooperation between members was reinforced by his request that members submit observations monthly and by issuing a large number of Circulars which allowed rapid feedback to observers (1); at least 79 circulars were issued. He also maintained a prodigious private correspondence with Section members, providing advice and encouragement, which led to one person commenting "in fact the amount of work he got through is almost incredible" (4). Not content with written forms of communication, he introduced a new idea to the BAA which is now commonplace: the idea of a section meeting where members could get together to discuss section matters and observing. Thus the first VSS meeting was held on 10 December 1906 at a London hotel, attended by 8 members. Such was the apparent success of this meeting that he suggested other sections take up the idea, remarking that one objective of such meetings is "that we want to infuse as much life into our work as possible" (67) .

Markwick recognised the need for observers to use standard comparison star sequences, which were based on Pickering's HCO photometry, and standard methods (e.g. Argelander's step method). Observers recorded the full estimate, including which comparison stars were used, an approach which is maintained in the VSS database even today. This allows the reduced magnitude of the variable to be recalculated should the comparison star sequence be updated later (68). He also required observers to record systematically the accurate time of observation, along with details of the instrument and magnification used, and an assessment of both the state of the sky and an estimation of the quality or "class" of the observation.

As a result of Markwick's enthusiastic and focused leadership (69), the VSS began to flourish and it soon became one of the most active sections, with a multiplicity of reports on its work being published in the Journal as well as the Memoirs (70). This in turn encouraged new members to join (71). As John Toone noted in his account of *British Variable Star Associations 1848-1908* (1), "Markwick deserves full credit for transforming the VSS from a minor club style reporting forum into a robust association of variable star observers, with a clear strategy and applying the most professional and exacting standards of the time". In his 1904 treatise on "The Observation of Variable Stars" (72), Markwick commented on the increasing interest in variables shown by professionals in the emerging branch of astrophysics:

"The study of variable stars is becoming more and more important, as the causes of which the light variations are due lie deep in the domain of cosmical physics, and in fact form some of the leading phenomena of the universe of stars..... Formerly the subject was practically left untouched in the program of the official observatories; and even now.........the bulk of the work is being done in private observatories or by amateurs without an observatory at all".

Markwick's comments could equally apply to much of variable star astronomy today. The approach and methods that he championed as VSS Director are essentially the same as those adopted by subsequent Directors up to and including the present time. Moreover, they were emulated by other variable star associations around the world.

Although Markwick relinquished the VSS Directorship in December 1909, handing the reigns over to Charles Lewis Brook (73) (1855 – 1939), he remained an active variable star observer for the remainder of his life and continued to publish analyses of VSS observations. His New Year greetings in 1924, a little more than a year before his death, read: "May we all have good weather, and health and strength to observe – and all unite in advancing out knowledge of stellar variation" (3).

**Nova Persei 1901 and the VSS nova patrol**

In his 1891 JBAA paper on *Naked-Eye Astronomy* Markwick wrote: (37) "supposing a new star of the third magnitude, or brighter, were to appear, how many amateurs are there who would be at once aware of the fact from personal knowledge? No doubt, in such a case, in the multitude of observers there is safety; someone or other would soon chance across it (leaving public observatories out of the question for the moment). At the same time, an amateur need be a pretty acute observer to note originally and independently the appearance of a new star of third magnitude; and yet, should some outburst of light occur in a far distant world, it is open to any to do it, if he is favoured with good weather at the right time."

It was only a decade later that the astronomical community was treated to a bight nova. In the early hours of the morning of 22 February 1901, the Edinburgh clergyman Thomas David Anderson (1853-1932) was about to retire for the night when he cast a final "casual" glance up to the sky and noticed a third magnitude star in Perseus. He reported his discovery to the observatories at Greenwich and Edinburgh. Several independent discoveries were reported later.

That evening, Markwick was entertaining a fellow officer at his residence in Devonport and astronomy was far from his thoughts: (74)

"a brother-officer, who was leaving the station next day prior to embarkation for India, was dining with me, and as it was the last evening we should be together after a service of over two years, during which we had been closely associated in duty, astronomy for the time gave place to hospitality. As we were smoking our cigars over the fire after dinner (the stars shining brightly outside—save the mark!) a telegram was handed to me from 'Gregg, The Lindens, St. Leonard's' "Just seen strange star left of Algol, 6.40 [GMAT; 18.40 UT]. What is it?"

Markwick immediately knew what he must do: (75)

"I asked my friend, who was acquainted with my astronomical proclivities, to allow me to investigate for a few moments, as this was a matter demanding serious attention. Stepping outside my front door, which faces west, I at once noticed a brilliant star above and to the left of Algol, which must be new. Here was a novelty for my friend, for you do not get an invitation every day to meet such a distinguished visitor as Nova Persei, particularly the first or second night of its arrival. Fair shone the moon and stars and we agreed at once to adjourn to the observatory, where we spent an hour or more observing the stranger."

Markwick's first estimate placed the nova at magnitude 0.72 and, later in the evening at 0.90. He sent a telegram back to Mr. Gregg (76), confirming that he too had observed the new star, and another to the Astronomer Royal. He then wrote up an account for the local paper "and my eldest son rode off on his bicycle to the office of the *Western Morning News*, and gave it to the sub-editor, who, I was informed, treated it in a somewhat nonchalant way, as quite an ordinary piece of news. However, it duly appeared next morning in that excellent journal....." (75) (77)

Nova Per 1901 (now known as GK Per (78)) was one of the brightest novae of modern times, reaching magnitude 0.2 at its brightest. Markwick and other members of the BAA VSS, as well as astronomers around the world, monitored the nova intensively. Markwick wrote several reports on the nova in the JBAA (79) and the *English Mechanic*, plus a BAA Memoir (80).

Some 3 years after Anderson's discovery of Nova Per, Markwick announced a "*Plan for watching the Region of the Milky Way for Novae*" under the auspices of the VSS (81). The Milky Way was divided into 6 sections and observers, which included Markwick and Gregg, were assigned to these areas with the aim of examining them regularly for novae. Charts were issued, but it appears that this project, in contrast to Markwick's other projects, never really took off.

Other novae that Markwick observed included Nova Aurigae 1892 (T Aur) (82) (Figure 15), which was also discovered by Anderson, Nova Geminorum 1912 (DN Gem) (83), Nova Aquilae 1918 (V603 Aql), the brightest nova of the 20$^{th}$ century (84) (85), and Nova Cygni III 1920 (V476 Cyg) (86). Markwick published a light curve of the latter nova in the November 1920 JBAA (Figure 16), based on his observations plus observations by W.F. Denning (1848-1931), who discovered the nova, and the observatories at Greenwich and Harvard College.

**Solar observations**

Although primarily known as a variable star observer, Markwick was also an active solar observer. He made his first solar observations on 19 July 1874, the day on which he was presented with the 2¾ inch refractor by his parents. A more intense interest in the sun was stimulated, as he recorded (20), by comments he read in the *English Mechanic* written by Arthur Cowper Ranyard. Over the years, he contributed many observations to the BAA Solar Section and in the last few years of his life he wrote a monthly column in the *English Mechanic* summarising recent solar activity. One of his favourite observations was of the "great" sunspot of January-February 1905 (87). He first noted the spot near the limb on the morning of 28 January. Over the next days, the group grew in complexity and he struggled to record all the "detail of penumbra and dark pores on paper, even with a finely pointed pencil" (see Figure 17). His final observation of the group was on 8 February as it approached the opposite limb of the sun, but by that time it was "dying away".

Markwick watched the not quite total solar eclipse on 17 April 1912 from Southsea Common (88). He sketched the curious, and somewhat amusing, appearance of his head's shadow on the ground at the time when the sun was a crescent (see Figure 18), noting (89) that the sketch:

"made at 11h. 58m. [11.58 UT] seemed to show whiskers on each side of the head, although I am quite free from such appendages. At this time these dusky fringes showed on both sides, and had their points, or curvature, turned upwards. The second sketch, made at 12h. 31m. [12.31 UT], or 33 minutes later, only showed fringes on the right-hand side of the head, and the points were now turned downwards. These changes were doubtless due to the varying position of the bright crescent of sunlight".

**The total solar eclipse of May 1900**

The track of the 28 May 1900 solar eclipse passed through the southern USA, across the Atlantic, reaching land in Portugal, then onwards through Spain, over the Mediterranean and into Algeria (90). Members of the BAA observed the eclipse from each of these countries and a report of their expeditions was published in a BAA *Memoir* (91) edited by E.W. Maunder (1851-1928). Markwick observed the eclipse from the deck of the Orient line's *SS Austral* (92) off Portugal and he contributed a chapter on his experiences to the *Memoir* (93). He evidently considered various options for viewing the eclipse, both land based and maritime, noting:

"My wife thought I was embarking on a 'wild goose chase', but even this form of sport *sometimes* results in success. So I made up my mind to risk it, as even if I saw nothing, a short sea trip after a long spell of worrying official duties would do one no harm, to say the least of it".

The *Austral* sailed from Plymouth, less than 1 hour from his residence, on 26 May. There was apparently only one other amateur astronomer on Board, Mr. W. Broadbent, of Huddersfield, who was also a fellow BAA member. Markwick struck up a friendship with the *Austral*'s "able commander", Capt. A.J. Coad (94). Markwick assisted Coad to navigate exactly to the centre line of totality. Regarding his eclipse observations, Markwick's "apparatus was simple; in addition to my eyes I had one piece of [smoked] glass......, a dark solar eyepiece cap...slightly smoked, one binocular magnifying 5 diameters with eyepieces smoked inside, another good binocular power about six, in natural clear state, a deck watch, 5s slow on GMT, note book and pencil. These were laid out on a travelling rug secured to one of the ship's seats on the open forward part of the promenade deck". A photograph of observers on the promenade deck is shown in Figure 19.

During the morning Markwick observed Venus with the naked eye in the clear blue sky and the advancing of the Moon across the Sun (Figure 20). At totality (Figure 21):

"Very roughly the Corona was like a band or ribbon of light, stretching from left of up to right of down, about the same width as the sun's diameter, with comparatively tiny aigrettes shooting out at the poles....There was a brilliant prominence, principally white, with a touch of pink on the lower limb...A wealth of detail was apparent all through the Corona, wisps and rays interlacing.....The general hue [of the Corona] was pure white or greyish white; the sky was blue all round the sun, and the effect of the silvery Corona projected on it, was beyond anyone to describe. I can only say it seemed to me what angels' wings will be like".

Only too soon the "sunlight flashed, and the exquisite vision passed – a vision which will be treasured up in the writer's memory as long as he is permitted to consider the heavens". (95)

Markwick disembarked the *Austral* in his old stomping ground of Gibraltar, where he visited some fellow Army officers. He also met with three other prominent astronomers who were returning from observing the eclipse (Figure 22): Norman Lockyer, Ralph Copland (1837-1905), Astronomer Royal for Scotland and the spectroscopist, Alfred Fowler (1868-1940) (96), before sailing back to England.

**Other observations**

Markwick observed a wide range of astronomical objects and phenomena in addition to variable stars and the sun. As we saw earlier, it was probably the appearance of a Comet that sparked his interest in astronomy and over the years he observed other bright comets including the "Great Comet" of 1882 (C/1882 R1). Figure 23 shows his drawing of Comet Perrine (C/1902 R1) in 1902.

As people eagerly awaited the return of Halley's comet in 1909, observatories around the world vied with each other to be the first to recover it as it headed towards the sun. Markwick, being very patriotic, was keen that the honour of recovery should fall to a British observatory. Although a photographic plate taken at the Royal Greenwich Observatory was later shown to have recorded the faint comet, the plate was not examined sufficiently early to announce the discovery, allowing Greenwich to be beaten to the discovery by Max Wolf on 12 September 1909 from Heidelberg in Germany. Markwick rued (97):

"Here we have the comet—or, rather, Greenwich has it—on a photographic plate; yet the discovery is announced subsequently to the taking of that plate by a Continental astronomer. So it was with Neptune; Cambridge Observatory had the planet on their charts, but through tardy examination the discovery was made on the Continent".

He hoped that "Halley's Comet, at its next first appearance, may be discovered at an Imperial observatory of the British Empire, situated, perhaps, in India, in the Soudan, in the centre of Australia, or the clear uplands of South Africa!" (97). In the event the next approach of Halley's comet was first detected by David Jewitt and G. Edward Danielson on 16 October 1982 using the 200 inch (5.1 m) Hale telescope at Palomar Mountain, USA. In spite of this, Markwick would no doubt have rejoiced that Jewitt was born in England!

Markwick had his first view of Halley's comet on the evening of 3 December 1909 (98): "it looked like a little dim patch of nebulosity, somewhat preceding the star 70 Tauri". Meanwhile a new bright comet had suddenly appeared in the evening sky, C/1910a (C/1910 A1), which the Markwick family glimpsed on the evening of 22 January 1922 (98):

"I, in company with my wife and daughter, began to watch for the new comet from an upper window of my residence [in Boscombe], which commands an uninterrupted view of the W.—S.W. sky. The weather was frosty, and although there was a dark stratum of haze and smoke due to numerous afternoon teas then in progress the sky, generally, was beautifully clear........ At 5h. 20m. [GMAT; 17.20 UT], Miss Markwick first 'spotted' the comet with a small pair of opera glasses, thus gaining the (domestic) comet prize. I followed suit in a few seconds, catching it in a prismatic binocular magnifying eight times".

He followed the comet through late January and early February. He attempted some spectroscopic observations, but his telescope didn't have sufficient light grasp to show anything definitively (99): "With the Hilger spectroscope (slitless) the spectrum was very dim; but there was a bright condensation, possibly in the green. I suspected also some dark bands or gaps. There was a background, or continuous spectrum, also; but the whole thing was so faint, it is difficult to describe exactly what was seen".

Apart from comets, Markwick also wrote numerous descriptions of his observations of the zodiacal light (100), meteors (101), lunar eclipses (102), planetary conjunctions (103) and an occultation of Saturn (104), amongst many other things. As is often the case with amateur astronomers then and now, he had an active interest in meteorology, writing a series of articles in the *English Mechanic* on the subject (105).He also wrote extensively about atmospheric phenomena such as unusual sunsets, sun pillars and a lunar rainbow (106). He also realised the negative effect that light and air pollution was having on astronomical observations from Britain's rapidly growing towns, such as his native Bournemouth, which led him to be an early advocate of the control of light and air pollution (107).

**Non-astronomical interests**

As an educated gentlemen of the Victorian and Edwardian age with an enquiring mind, and one who apparently had time on his hands during his various Army postings, whilst on leave and during retirement, Markwick maintained a broad range of personal interests outside astronomy. He wrote many letters to the *English Mechanic* in connexion with his interests in violins (108), antiquities (109), the genealogy of the Markwick family (110), aviation (111) and railway travel (112).

There were other curiosities of a highly practical nature. For example in a letter under the title *Curing Sea Fish* he asked (113) "Can anyone kindly give me any hints on the salting and curing of saltwater fish or refer me to any literature on the subject?", although no reason is given for his interest in the subject. In another letter he asked (114) "Will any reader kindly give particulars for the different materials that could be used for making smoke-rockets for testing drains". This was in September 1914, about the time of his move to Dublin, and one could speculate that perhaps there were sanitation problems in his new residence. An example of Markwick's wide-ranging experience which he wished to share with readers of the *English Mechanic* was a letter he wrote about choosing the correct design of waterproof coats for different climates (115):

"I do know from experience that there are two classes of these garments: one adapted for foreign climates, and one not adapted. I had one of the latter, regulation pattern, which was right enough in the United Kingdom; but the Gibraltar "Levanter" (a hot wind loaded with moisture) very soon played "Old Harry" with the garment in question, so that it could very easily be "peeled" to pieces, if I may be pardoned the slang expressions. Then, in South Africa, I remember another garment, not adapted for foreign climates, which became so stiff and rigid that it could with a little balancing be made to stand upright by itself. I now have a waterproof guaranteed for all climates, which so far wears excellently".

Markwick was a great reader of the classics of English literature, both novels and poems, and he often quoted from them in his writings. He wrote treatises on the astronomical events and allusions mentioned in the poems of Milton (116), Wordsworth (117) and Alexander Pope (118) and the works of Shakespeare (119).

**The final years**

Having followed his patriotic duty and re-enlisted in the Army at the outbreak of the First World War, Markwick remained in Ireland until well after the end of the war. Finally, upon demobilisation in 1919 he set up home with his wife in West Moors, Dorset, just north of Bournemouth. He did most of his variable star observing at his new residence with the 8½ inch (21.6 cm) Calver reflector (5) (120). There he also engaged a local carpenter to build a small observatory in the "Romsey style" that was fashionable at the time (121). It had a wooden frame and was covered in rubberised canvas, housing his 4 inch (10 cm) Grubb refractor which he mainly used for solar observation. His friends nick-named the observatory "the beehive" and in his correspondence in later years he gave his address as "The Beehive Observatory, West Moors".

Markwick fell ill in early 1925 (122) and his longstanding friend Arthur Mee (123) (1860-1926; Figure 24) reported (124) in the 8 May 1925 edition of the *English Mechanic*: "There will be widespread regret at Colonel Markwick's illness, necessitating the discontinuance (let us hope for a time only) of his valuable solar reports". In the event, Markwick's last publication was one of his monthly solar reports: that for January 1925, published in the *English Mechanic* on 27 February (125). He passed away on 4 July 1925, two weeks before his 72$^{nd}$ birthday. His *English Mechanic* obituary, written by W. Strachan (126), summarised his willingness to help his fellow astronomers (5):

"But it was the Astronomical moiety of his psychology that appealed most strongly, of course, to readers of "Ours" (127) and astronomers the world over—but especially to amateur astronomers. Markwick was an observer of great ability and experience, and his vast fund of practical knowledge in this respect was ever at the disposal of the veriest tyro in telescopic work. He was never so happy as when 'helping' astronomical "lame dogs over stiles" and this assistance and guidance were always given with an ability, charm of manner, and kindliness peculiarly his own...... As "skipper" of the variable star section of the Association [i.e. the BAA] over a long period, he proved himself a great and real Director—painstaking and able to the last degree..... Col. Markwick was a fine soldier, a very distinguished astronomer, a staunch friend, and true English gentleman. Nor will his many varied and valuable services to astronomy do otherwise than remain a cherished memory, especially with the amateur "watcher of the skies" for long years to come".

The following March, Markwick's wife put his telescopes and observatory up for sale via a private advertisement in the *English Mechanic* (128) (Figure 25). Presumably these were sold fairly swiftly as they were not advertised further and in June Mrs. Markwick entered another advertisement for a solar diagonal (129). The 8½ inch (21.6 cm) Calver is current located in the visitors' centre at the Auckland Observatory in New Zealand (Figure 26), where it has recently been renovated. The back of the mirror is etched: "Made by G. Calver for E. E. Markwick, 1890". It was donated to Auckland Observatory in 1969 by Mr. Ernie Curtis, a long-standing member of the Auckland Astronomical Society, who had bought the telescope in 1957 (130) .

**Coda**

The Victorian era was a period of great prosperity for the British people, characterised by a long period of peace, known as the *Pax Britannica*, and economic, colonial, and industrial consolidation. Cecil Rhodes (1853-1902) is purported to have claimed "to be born English is to win first prize in the lottery of life" (131), reflecting the good fortune that many Britons enjoyed, at least those towards the top of the socio-economic ladder. Markwick, who was born in the same year as Rhodes, leading the life of an Army officer both in the mother country and in the colonies, may be viewed as one of the products of this unprecedented period of wealth, stability and national confidence. Moreover he possessed the education, self-assuredness and sheer curiosity of many a Victorian gentleman, along with the free time available to pursue a range of interests, including astronomy. This curiosity was also shared by the explorers of the era, such as Richard Francis Burton (1821-1890) and David Livingstone (1813-1873), who set out to find new lands and new peoples. The astronomical historian, Richard Baum (132), has likened the astronomers of the Victorian age to these explorers, pushing back the frontiers of their science and opening up new areas of understanding. In the same way Markwick can, therefore, be described as an explorer of the skies opening up new frontiers. Rather than sailing up new rivers or travelling across mountain ranges to make discoveries, Markwick navigated the night skies with his modest instruments, eventually leading him to discover new variable stars. He possessed a strength of character, even doggedness, characteristic of explorers, which enabled him to continue the painstaking search for new objects over many years. Because the study of variable stars by amateurs is nowadays relatively common, one might underestimate his pioneering efforts to open up this adventure to others who wished to join him.

Markwick's leadership and organisational skills, which were no doubt developed during an illustrious military career, made him a natural choice as President of the BAA. But he possessed other personal qualities: he was intelligent, affable, sociable, good humoured and witty. He was respected and liked by those who came into contact with him. He went out of his way to help and encourage others and no doubt attracted many people to the hobby of astronomy. These qualities also led him to become one of the most successful section Directors of all time. He reorganised the VSS, establishing standard methods and encouraging cooperation between observers by introducing a list of target stars and comparisons. He provided regular encouragement and feedback to the band of observers, thereby allowing them to maintain their enthusiasm, and he analysed their combined results, writing papers which would make the fruits of their labours available to other astronomers around the world and in generations to come. In this way he set the VSS on a path to become in a short space of time one of the most respected variable star organisations, a path along which it continues to this day. His VSS became a model that other variable star organisations around the world would emulate.

**Acknowledgements**

I am most grateful for the assistance I have received from many people during the research for this paper. Eric Hutton prepared an extensive extract of items from the *English Mechanic* relating to Markwick. Mike Saladyga (AAVSO) provided copies of Markwick's manuscript observations of variable stars from the AAVSO archives. Roger Pickard provided details of Markwick's BAA VSS observations from the VSS database. Bryan Stokes (School Archivist, King's College School, Wimbledon) searched the King's College School archives and provided information on Markwick's school days. Sheridan Williams has saved me a great deal of time I would otherwise have spent searching through library copies of the JBAA by diligently scanning the Journals and making them available on line to Members (a truly wonderful resource) – and he fast-tracked the availability of some specific editions to assist my research. Richard Baum provided much encouragement and inspiration to pursue this research and I have truly valued the fascinating discussions I have had with him on matters of astronomical history. Peter Hingley (RAS Librarian) looked after me on numerous visits to the RAS library and provided details of Markwick's RAS Fellowship application, as well as retrieving lesser known publications to which Markwick contributed such as the *Cambrian Natural Observer*. Tom Williams (Rice University, Texas and AAVSO) provided information about Markwick's correspondence with E.C. Pickering. Grant Christie (Auckland Observatory) generously took photographs of Markwick's Calver reflector and provided information about its history. Bryn Jones kindly provided the photograph of Arthur Mee. This research made use of the NASA/Smithsonian Astrophysics Data System. Finally, I thank my referees for their helpful and constructive comments before swine," it refers to a quality unbefitting those who partake of it. In this case, the "general" are the multitude to which Hamlet is referring.

16. Address: Innisfallen, Campbell Road, Boscombe.

17. Supplement to the London Gazette, 1 January 1919.

18. Markwick's address was The Knowle, West Moors, Dorset.

19. Amy was born in Rochester, Kent, UK in 1863 according to the 1911 census.

20. Markwick E.E., JBAA, 2, 496 (1892). Markwick's paper on the "Evolution of an amateur astronomer" describes his emerging interest in astronomy.

21. In his 1892 JBAA paper on the "Evolution of an amateur astronomer", Markwick recalls reading these articles, probably by Herschel, in the magazine "Good Words", which were published "over 20 years ago". He goes on to say that they "evoked a very deep interest at the time in astronomy generally, being written in a very taking style"

22. Airy delivered a series of six lectures in 1848 on 'Popular Astronomy' to support the founding of the Ipswich Mechanics Institute. They were subsequently published

23. One of the earliest of Proctor's books that he read was "Saturn and its System", Proctor R. A., Longman and Green, London (1865).

24. As discussed later, the first observation with his new telescope was of the sun. More than 30 years later, in 1905, he wrote: "There are not many toys that last as long as this, except it be a violin!" [Markwick E.E., JBAA, 15, 230 (1905)].

25. Described on many occasions by Sir Patrick Moore "as about as steady as a blancmange". For example: Moore P, The Amateur Astronomer, 7th edition, publ. Lutterworth Press (1971).

26. Markwick E.E., English Mechanic, 1809, 339 (1899).

27. Markwick had a copy of an 1873 edition of Webb's book.

28. Markwick E.E., English Mechanic, 1867, 472 (1901).

29. Although he spent the majority of time in Boscombe, the 1911 Census shows him resident as householder with his wife and daughter at 14 Devonshire Hill, Hampstead, London.

30. Report of the BAA meeting of 29 October 1913: JBAA, 24, 1 (1913).

31. According to Peter Hingley, it was common at the time for several people to sign the application form (in Markwick's case 4 people) even if they did not personally know the candidate.

32. In January 1879, Sadler wrote a strongly worded criticism of Admiral Smyth's "Cycle of Celestial Objects", questioning the authenticity of Smyth's observations. The RAS Council subsequently expressed their regret at having published Sadler's comments.

objective prism to identify stars with unusual spectra, many of which turned out to be variables. This approach rapidly and greatly increased the number of known variables, most of which had hitherto been discovered visually.

52. Pickering E.C., ApJ, 4, 142 (1896).

53. Markwick E.E., English Mechanic, 1720, 83 (1898).

54. Markwick E.E., JBAA, 5, 247 (1895).

55. Markwick E.E., AN, 138, 213 (1895).

56. Markwick E.E., English Mechanic, 1532, letter 36462 (1894).

57. Pickering E.C., HCO Circular 4, 1 (1895).

58. Samus N.N. et al., General Catalogue of Variable Stars (GCVS database Version 2011 Jan), available at http://www.sai.msu.su/gcvs/gcvs/.

59. Markwick E.E., AN, 140, 93 (1896).

60. Pickering E.C., AN, 135, 161 (1894).

61. FitzGerald A.P., IrAJ, 7, 213-219 (1966). Gore's RAS obituary is in MNRAS, 71, 256-257 (1911).

62. Gore J.E., Mem. Brit. Astron. Assoc., 3, 2, Variable Stars (1894).

63. Gore J.E., Mem. Brit. Astron. Assoc., 5, 2, Variable Stars (1896).

64. As Mike Saladyga and Tom Williams say (Advancing Variable Star Astronomy, publ. Cambridge University Press, 2011) it wasn't until as late as 1906 that E.C. Pickering noted in the publication of his "A Durchmusterung of Variable Stars", that the value of visual observations of variable stars was no longer centred on the individual's attention to a single star, but on the recruitment of many observers who were able to observe as many stars as possible. Pickering's view had been stimulated by the rapidly, increasing numbers of known or suspected variable stars as a result of photographic all-sky surveys, including HCOs own survey.

65. Report of the BAA meeting of 27 December 1899; English Mechanic, 1809, 339 (1900).

66. Markwick E.E., JBAA, 10, 112 (1900).

67. Markwick E.E., JBAA, 17, 132 (1907).

68. Early comparison sequences were sometimes unreliable and were often obtained by visual estimation. Over the years, sequences have been improved by new photometry. Since the BAA VSS records the full estimate, the magnitude of the variable can easily be re-reduced at any future time using the new sequence. Most other variable star associations only record the reduced magnitude of the variable, rendering updates based on new sequence photometry difficult or impossible.

69. Markwick wrote frequently on the subject of variable stars in the English Mechanic. Notably, he wrote a 12-part series of articles on the nature and observation of variable stars which did much to popularise the subject. Part 1 appeared on 23 November 1906 [Markwick E.E., English Mechanic, 2174 (1906)] and part 12 on 3 May 1907 [Markwick E.E., English Mechanic, 2197 (1907)].

70. Markwick published two BAA Memoirs summarising the VSS work on long period variables between 1900-1904 (Markwick E.E., Mem. Brit. Astron. Assoc., 15 (1906)) and 1905-1909 (Markwick E.E., Mem. Brit. Astron. Assoc., 18 (1912)). Markwick introduced a standard reporting format in these Memoirs which was continued by subsequent VSS Directors, Charles Lewis Brook and Felix de Roy. Marwick acknowledged that Brook had also made a significant contribution to compiling the lists of observations published in the Memoirs from the original observations submitted by members.

71. The Memoir covering the years 1900-1904 [Markwick E.E., Mem. Brit. Astron. Assoc., 15, (1906)] contains observations by 21 members of the Section and the Memoir covering the years 1905-1909 [Markwick E.E., Mem. Brit. Astron. Assoc., 18, (1912)] has observations by 25 members.

72. Markwick E.E., Popular Astronomy, 12, 193 (1904).

73. Brook's life and VSS Directorship are described in: Shears J., JBAA, accepted for publication (2011). The reader is also referred to two obituaries: Lindley W.M., JBAA, 50, 33 (1939) and Maunder A.S.D., MNRAS, 100, 233 (1940).

74. Markwick E.E., English Mechanic, 1875, 54 (1901).

75. Markwick E.E., JBAA, 11, 104 (1901).

76. Mr. Ivo F.H. Carr Gregg. Carr's own report of his independent discovery of Nova Persei appeared in: Carr I.F.H.C., JBAA, 11, 197 (1901).

77. In additional to detailed descriptions of the events of the evening in the JBAA and the English Mechanic already cited, Markwick also contributed a report to the journal of the Astronomical Society of Wales: Markwick E.E., Cambrian Natural Observer, 4, 93 (1901).

78. GK Per is monitored by BAA VSS members, and others around the world, to this day. From time to time it is observed to undergo dwarf nova type outbursts.

79. Markwick E.E., JBAA, 11, 242 (1901); Markwick E.E., JBAA, 12, 20 (1901).

80. Markwick E.E., Mem. Brit. Astron. Assoc., 10, number 3, "Nova Persei 1901" (1902).

81. Markwick E.E., JBAA, 14, 122 (1904).

82. Markwick E.E., English Mechanic, 1410, 127 (1892). Although Markwick does not list the observers' names in his letter, he says he gleaned the data from other reports submitted to the newspaper.

83. Markwick E.E., JBAA, 22, 285 (1912).

84. Brook C.L., JBAA, 28, 239 (1918).

85. During World War One, according to the VSS database, Markwick made very few variable star observations (typically 20 to 30 per year, compared to several hundred in the years before and after). The exceptional nature of Nova Aquilae presumably attracted his attention.

86. Markwick E.E., JBAA, 31, 82 (1920).

87. Markwick E.E., JBAA, 15, 230 (1905).

88. This was an unusual hybrid eclipse, or annular/total eclipse, which was seen as an annular eclipse along parts of the track and as total on others. The eclipse was only partial in England.

89. Markwick E.E., English Mechanic, 2457, 295 (1912).

90. The eclipse path went on into Libya, then part of the Ottoman Empire, but here the eclipse was becoming very short.

91. Maunder E.W., The Total Solar Eclipse 1900. Report of the expeditions organised by the British Astronomical Association to observe the total solar eclipse of 1900 May 28, publ. BAA (1901).

92. The Austral was the largest and most luxurious of the Orient Line steamers, with a gross tonnage of 5,588 and a 1,000 horsepower engine. She was launched in 1881, but the following year sunk at her coaling berth in Sydney Harbour. The next year she was refloated and refitted. She made many voyages between the UK and Australia.

93. Markwick E.E., Chapter 3: "At Sea" in Maunder E.W., The Total Solar Eclipse 1900, publ. BAA (1901).

94. In March 1917, Coad was captaining the TSS Orsova when it was torpedoed by a German U-boat just off the Eddystone Light. The port side of her engine room was holed, killing six of the duty crew, but Coad managed to beach her in Cawsand Bay inside Plymouth Sound.

95. During the evening after the eclipse, Markwick attempted to ascertain the time of the evening when the sky brightness equalled that during totality. He asked several passengers for their opinion, but there was a large diversity of view which rendered the experiment unsuccessful. See: Markwick E.E., English Mechanic, 1839, 402 (1900).

96. Fowler, a member of Lockyer's 1900 eclipse expedition party, was an astrophysicist at Imperial College London. He was an expert in spectroscopy, being one of the first to determine that the temperature of sunspots was cooler than that of surrounding regions. He was elected FRS in 1910.

97. Markwick E.E., English Mechanic, 2323, 210 (1909).

98. Markwick E.E., English Mechanic, 2340, 610 (1910).

99. Markwick E.E., English Mechanic, 2342, 33 (1910).

100. For example, he describes the zodiacal light from Cape Town in: Markwick E.E., English Mechanic, 1281, 137 (1889).

101. Markwick writes about the 1898 Leonids in: Markwick E.E., English Mechanic, 1757, 349 (1898).

| Star | GCVS classification (58) | GCVS mag. range (58) |
|---|---|---|
| U Cep | Eclipsing binary (EA) | 6.75 - 9.24 V |
| U Ori | Mira | 4.8 - 13.0 V |
| R Leo | Mira | 4.4 - 11.3 V |
| R UMa | Mira | 6.5 - 13.7 V |
| S UMa | Mira | 7.1 - 12.7 V |
| S Vir | Mira | 6.3 - 13.2 V |
| β Per (Algol) | Eclipsing binary (EA) | 2.12 - 3.39 V |
| α Ori (Betelegeuse) | Semi-regular (SRC) | 0.0 - 1.3 V |
| T Mon | Cepheid | 5.58 - 6.62 V |
| R CrB | R CrB | 5.71 - 14.8 V |
| X Her | Semi-regular (SRB) | 7.5 - 8.6 p |
| β Lyr | Eclipsing binary (EB) | 3.25 - 4.36 V |

Table 1: Markwick's first list of variable stars to be observed by VSS members

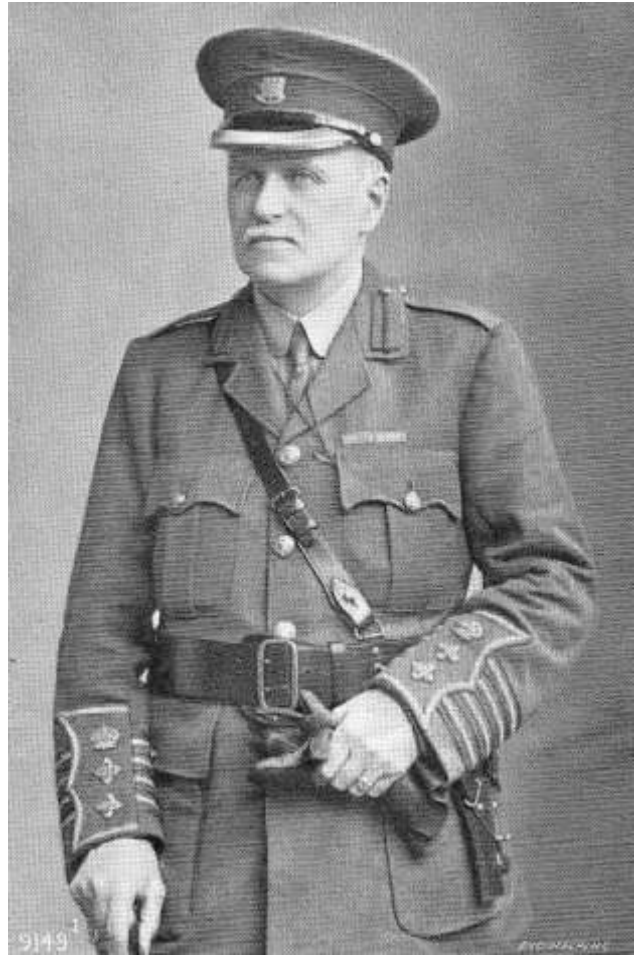

Figure 1: Col. E.E. Markwick, CB, CBE, FRAS (1853 – 1925)

BAA Presidential portrait

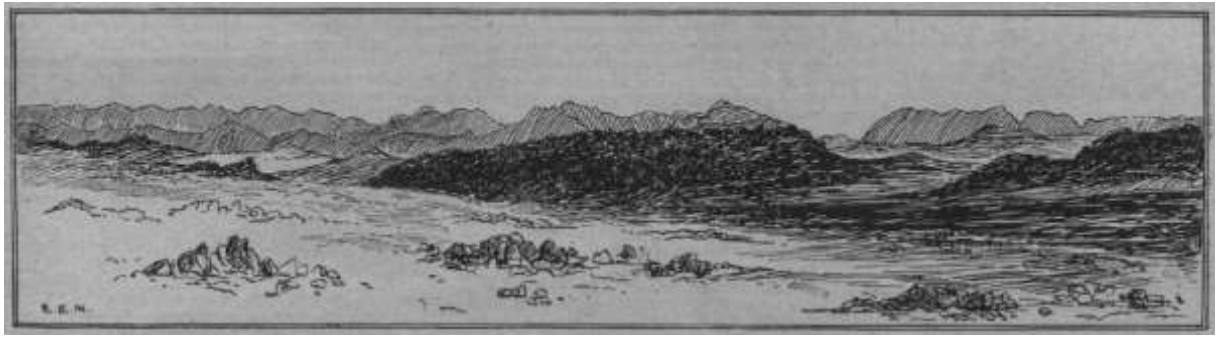

Figure 2: View of Magaliesberg Range, near Pretoria (133)

(E.E. Markwick)

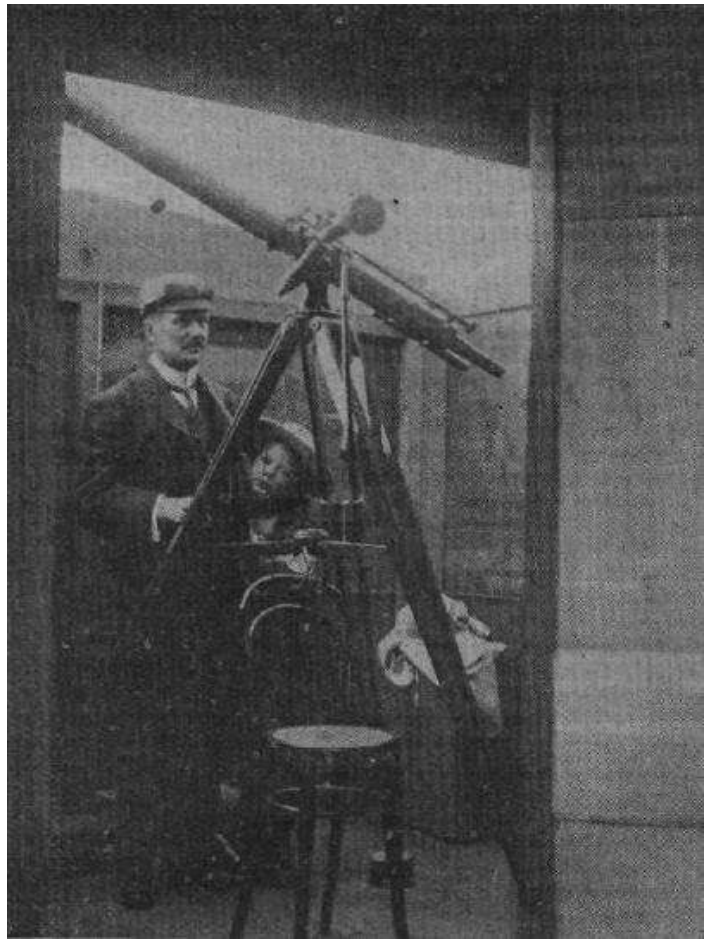

Figure 3: Markwick, his daughter and the 2¾ inch refractor in the Devonport observatory (134)

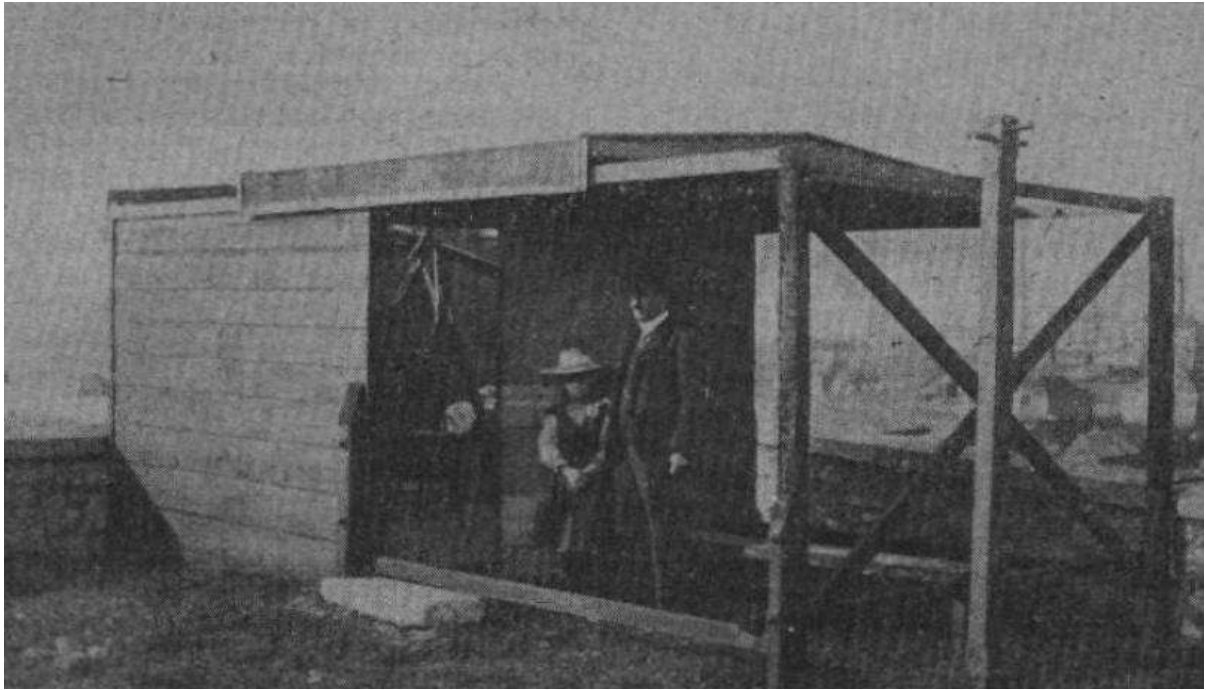

Figure 4: Markwick's observatory at Devonport, ca. 1899 (134)

Markwick and his daughter are standing next to the observatory. The photograph was taken by one of his sons. Markwick noted: "The whole was put up to my general design by someone a little more skilled than the oft-quoted "village carpenter", but the latter could no doubt put up just as good a house under proper supervision". The same photograph also appears in the Cambrian Natural Observer (135)

Figure 5: Markwick's RAS Fellowship application form

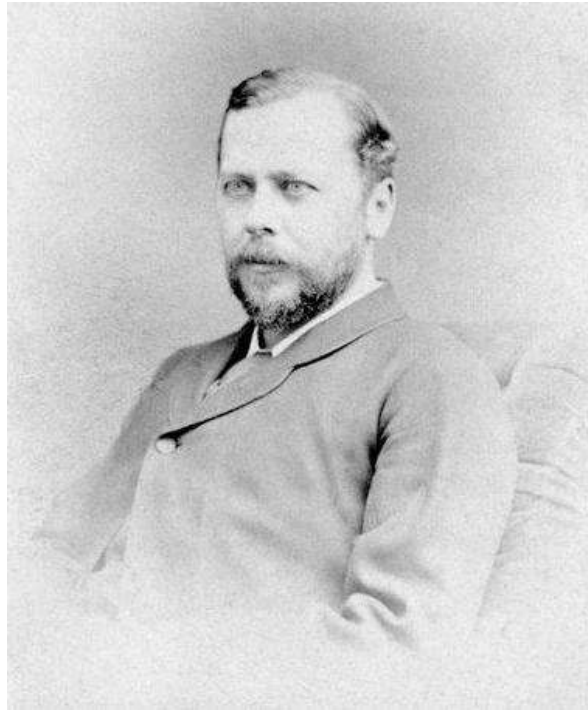

Figure 6: A. Cowper Ranyard (1845-1894)

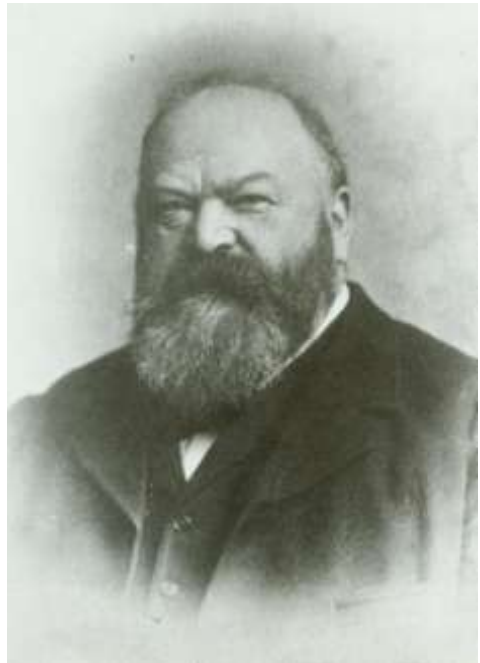

Figure 7: Andrew Ainslie Common (1841-1903)

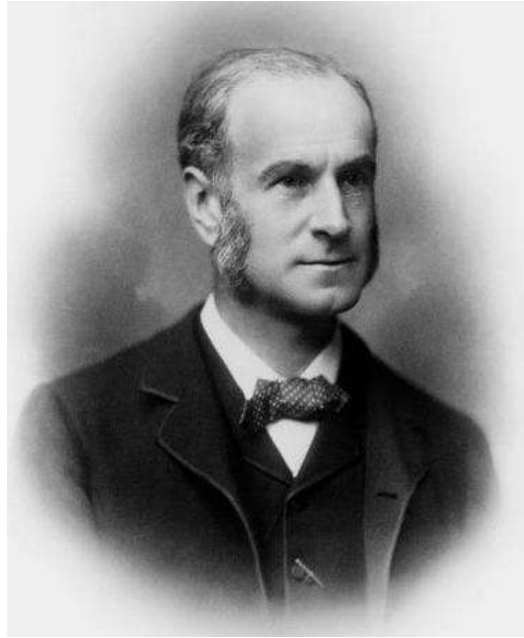

Figure 8: E. B. Knobel (1841-1930)

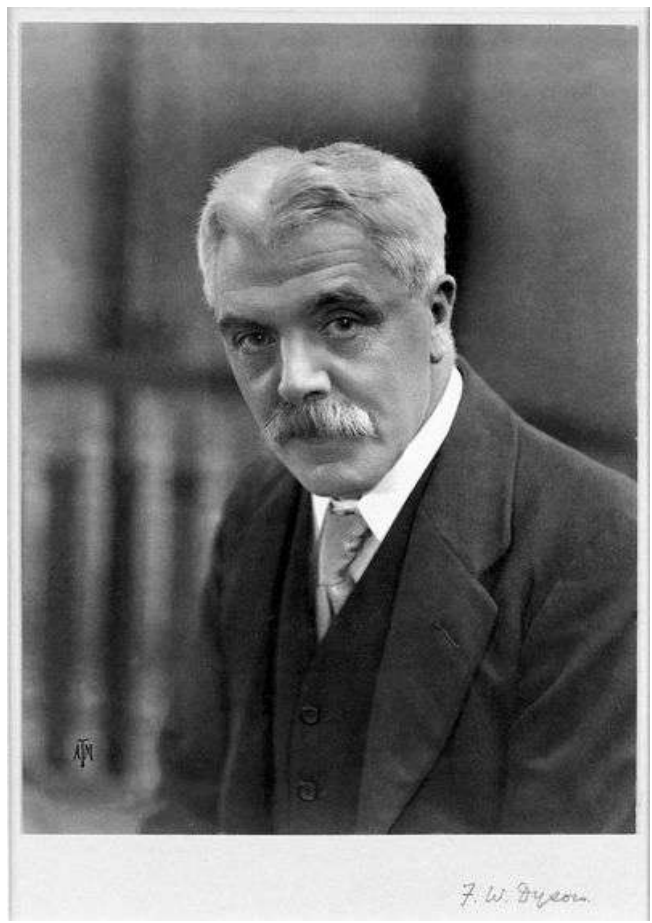

Figure 9: F.W. Dyson (1868-1939)

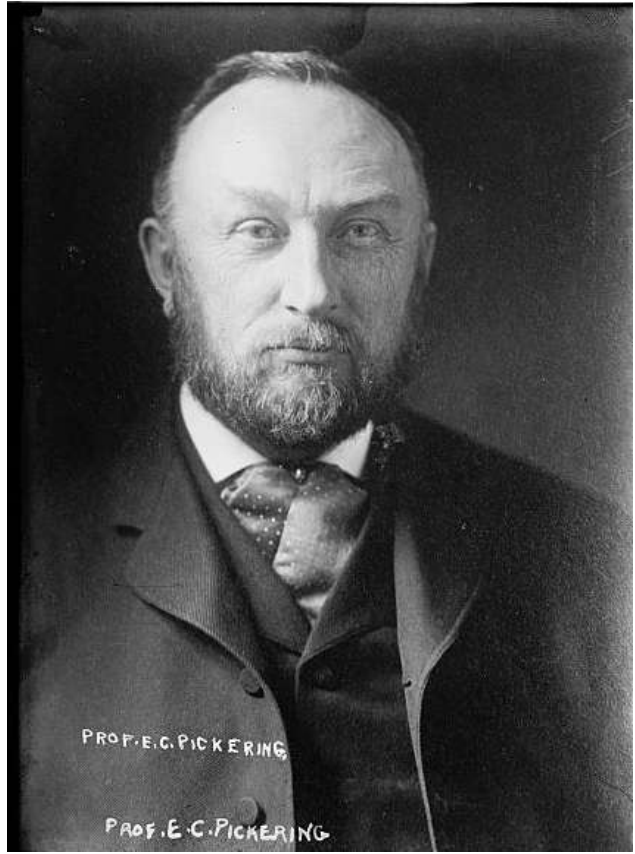

Figure 10: E.C. Pickering (1846-1919)

OBSERVATIONS of (Ch. 6202) u Herculis

By Col. E. E. Markwick

| Date. | Sky. | Instr. | Class. | Light Estimate. | Ded. Mag. | Remarks. |
|---|---|---|---|---|---|---|
| 1888 | | | | | | |
| June 13 | | | | a little > 69 Herculis; say 5m | 4.70 | Comparison stars |
| Aug 2 | | | | do | 4.70 | F 69 Herculis 4.50 H.R. |
| 13 | | | | do | 4.70 | · 72 — · 5.36 |
| Sept 2 | | | | Plainly < 69 | 5.2± | P — " — 4.14 ✓ |
| 5 | | | | do | 5.2± | The hour may be taken as correct within 5 min. |
| 28 | | | | do | 5.2± | Col. Sky. |
| Oct 9 | | | | do. Est 5m | 5.2± | 1 — v. good, good, or an average fair, working night, with no obstruction |
| 1889 | | | | | | |
| May 21 | | | | < P. about = 69 | 4.80 | 2 = Some obstruction such as cloud, haze &c. |
| 24 | | | | 3 steps < 69 | 5.10 | 3 very bad. |
| June 5 | | | | Plainly below 69 | <4.80 | T = Twilight M = Moonlight |
| July 21 | | | | do | <4.80 | "Instr." Col. |
| Sept. 14 | M | | | <69 | <4.80 | B = Binocular N.E. = Naked eye. |
| 21 | | | | about = 69, but usual >72 | 4.80 | Col. Class. |
| 1890 | | | | | | |
| July 14 | | | | <69 | <4.80 | 1 — a good reliable observer |
| Aug. 3 | M | | | slightly <69 | .75:1 | 2 — not as good, slight doubt, &c |
| Oct 10 | | | | about = 69. ½ or ¾ m. >72 | 4.76 | 3. very poor, shaky, unreliable |
| Nov 9 | | | | slightly <69. nearly a whole mag. >72 | 4.68 | "Light Estimate," generally in steps — 1 step being supposed = 0.1 mag |
| 1891 | | | | | | |
| April 26 | | B | | thought slightly >69 ½ m. >72 | 4.78 | > = brighter than |
| May 8 | | | | nearly a mag > 72. a little <69 | 4.68 | < = fainter than |
| July 15 | | | | a whole mag >72. =69 | 4.58 | |
| 29 | | | | do do do | 4.58 | |
| Aug 3 | | | | a little <69 | 4.90 | |
| Oct 22 | | B | | Decidedly <69, but this was not so apparent with N.E. ¾ m >72 | 5.09 | Continued |

Figure 11: Some of Markwick's observations of u Her, made between 1888 and 1891 (136)

*Accepted for publication in the Journal of the British Astronomical Association*

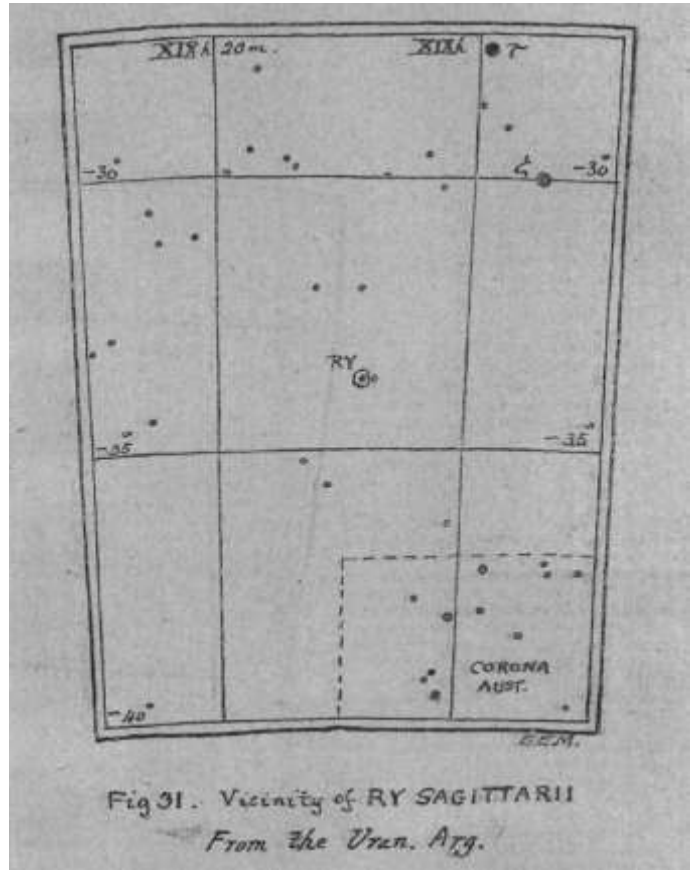

Figure 12: Markwick's chart of the vicinity of RY Sgr (48)

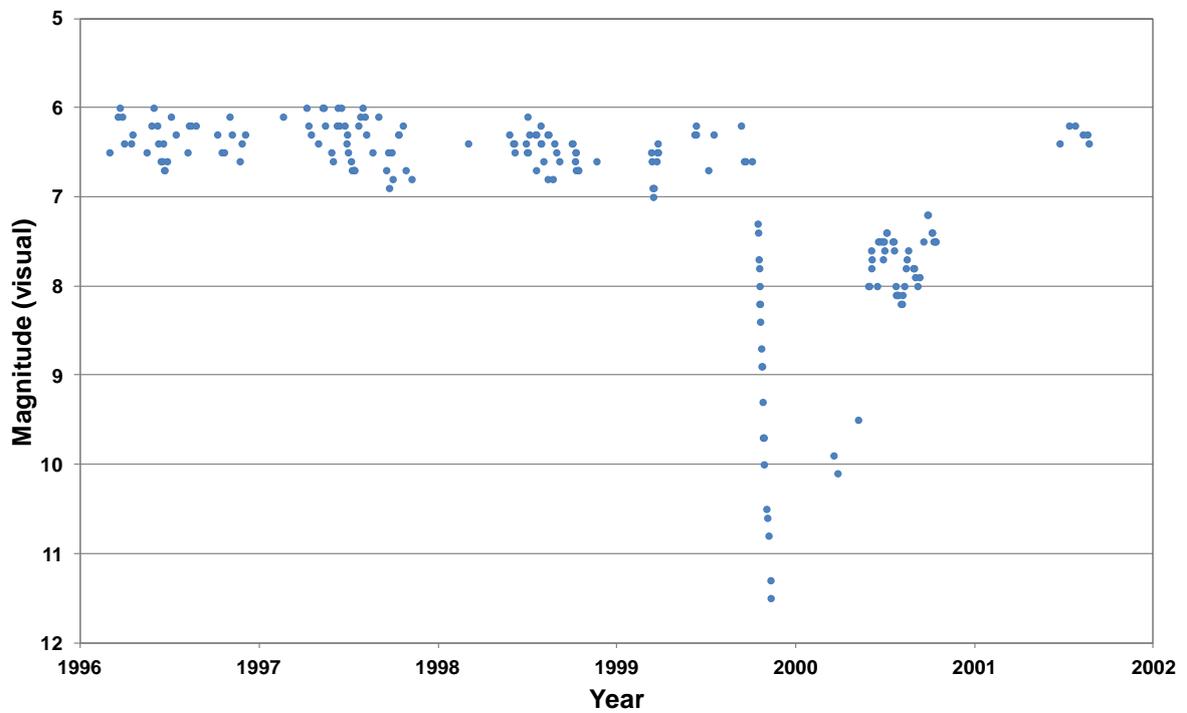

Figure 13: Light curve of RY Sgr between 1996 and 2002
Data from BAA VSS

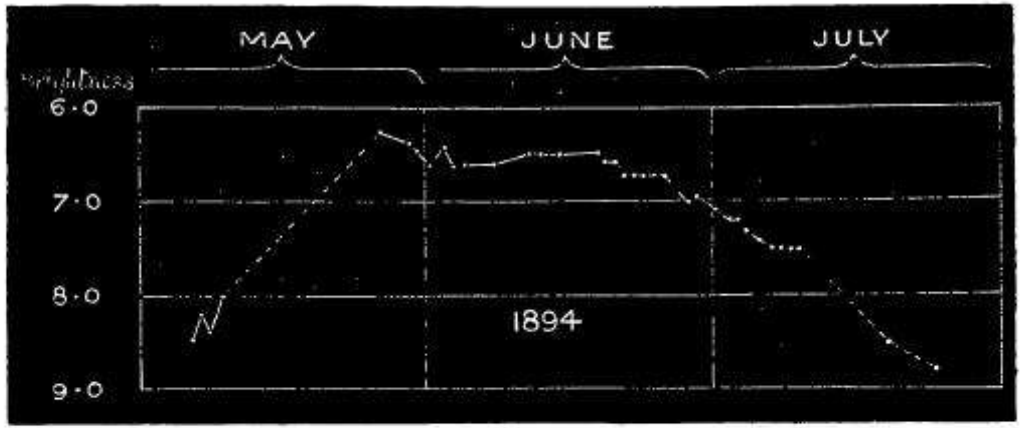

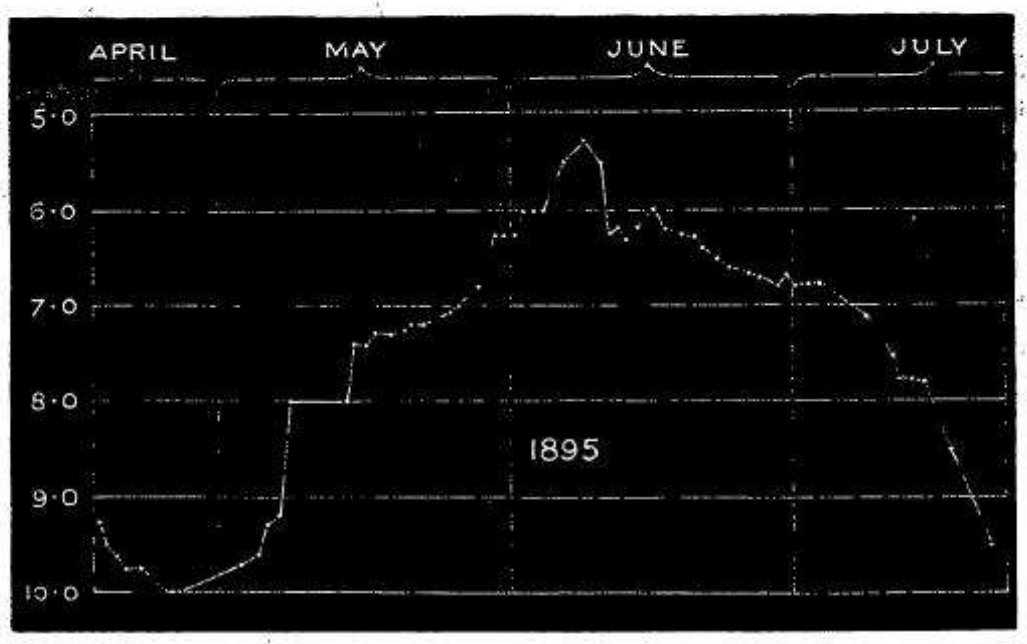

Figure 14: T Cen in 1894 and 1895
Markwick's plots of his own observations from reference (137)

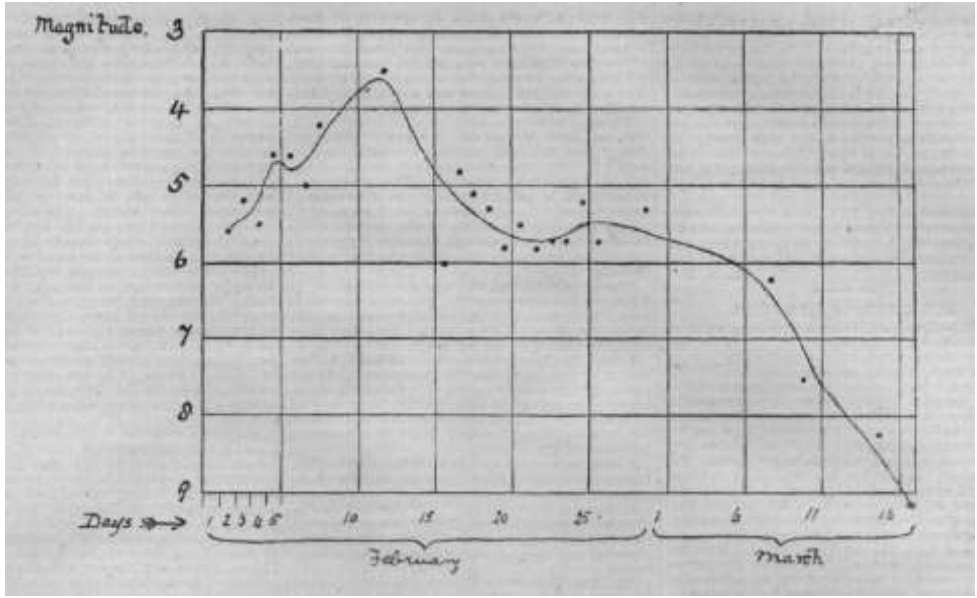

Figure 15: Markwick's light curve of Nova Aur 1892

Markwick plotted observations from February and March 1892 contributed by a number of observers to the *English Mechanic* (82)

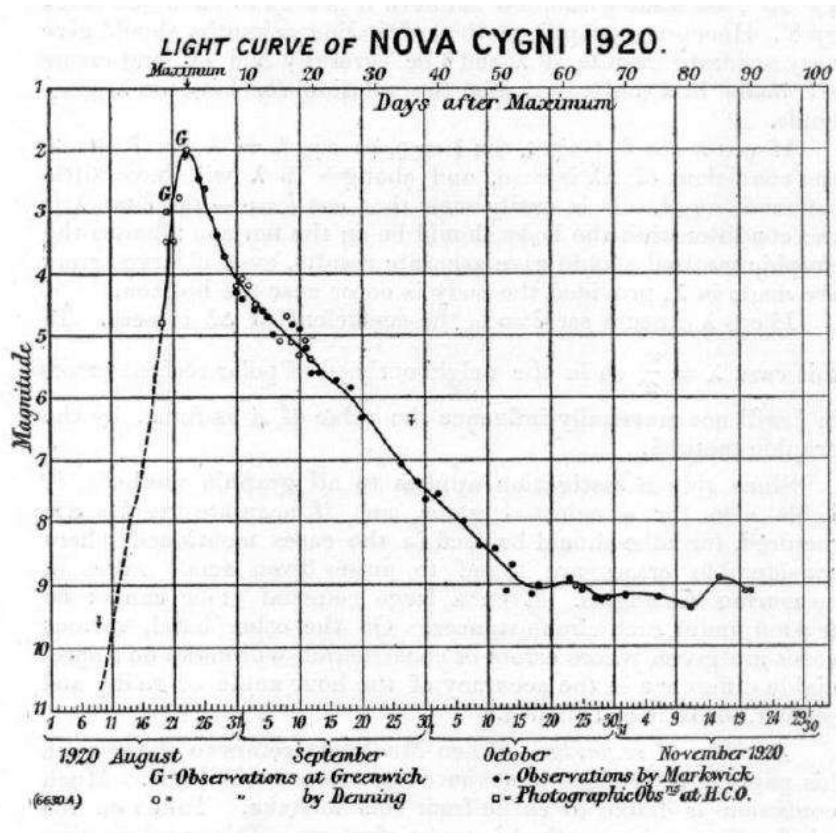

Figure 16: Markwick's light curve of Nova Cyg 1920

From reference (86)

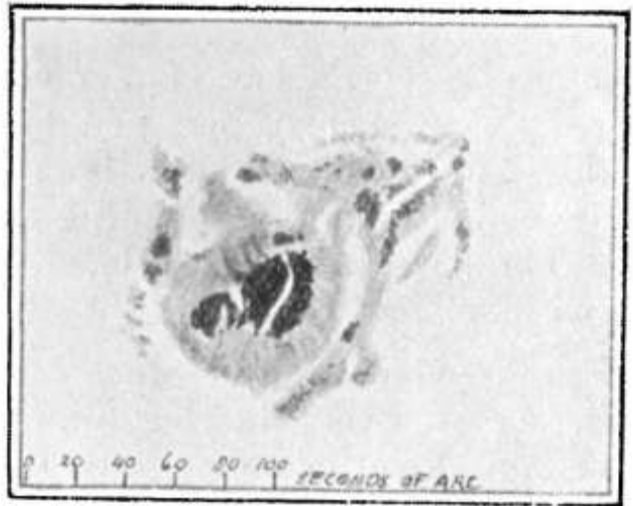

Jan. 31, 9ʰ 55ᵐ.

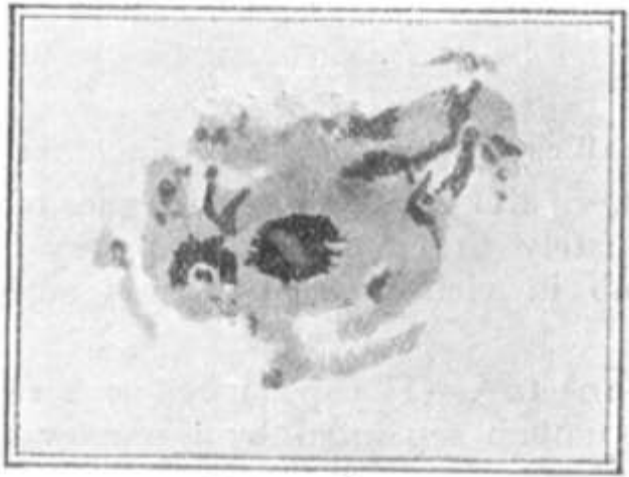

Feb. 1, 0ʰ 20ᵐ.

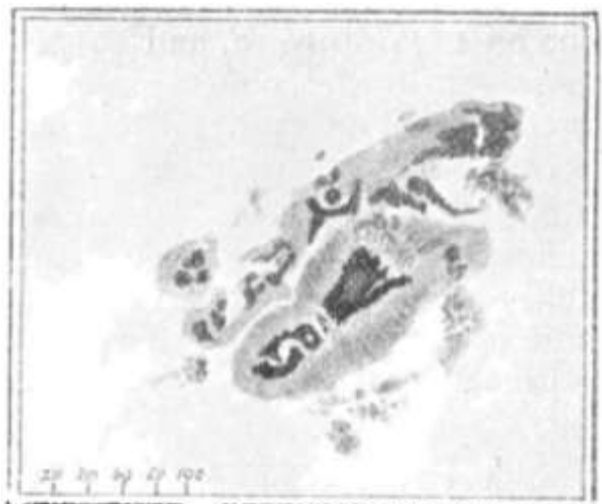

Feb. 3, 10ʰ 20ᵐ.

Figure 17: Markwick's drawings of the "great" sunspot of January-February 1905 (87). Times are UT

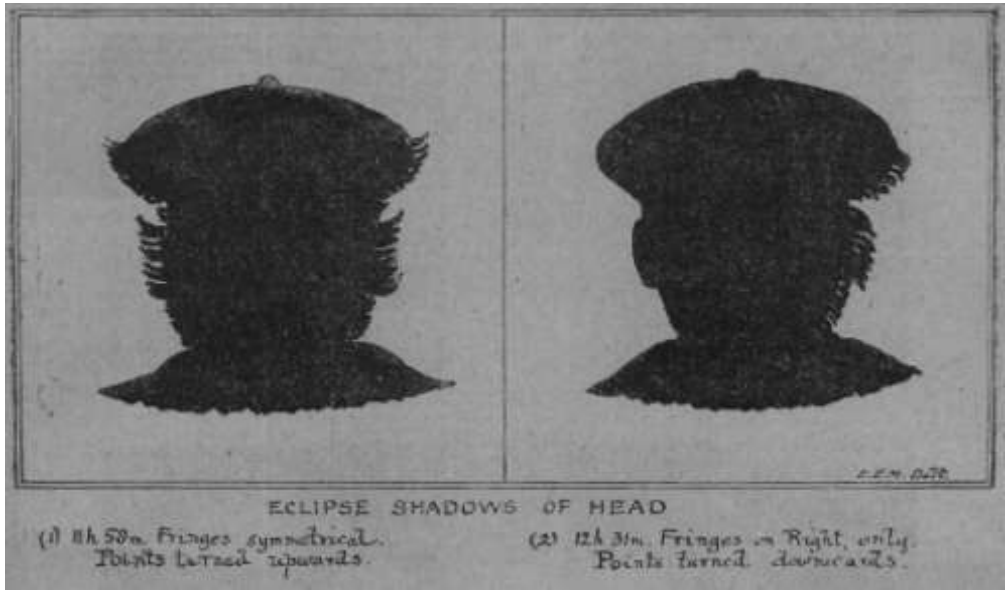

Figure 18: Sketches of the shadow cast by Markwick's head during the eclipse of 17 April 1912 (89)

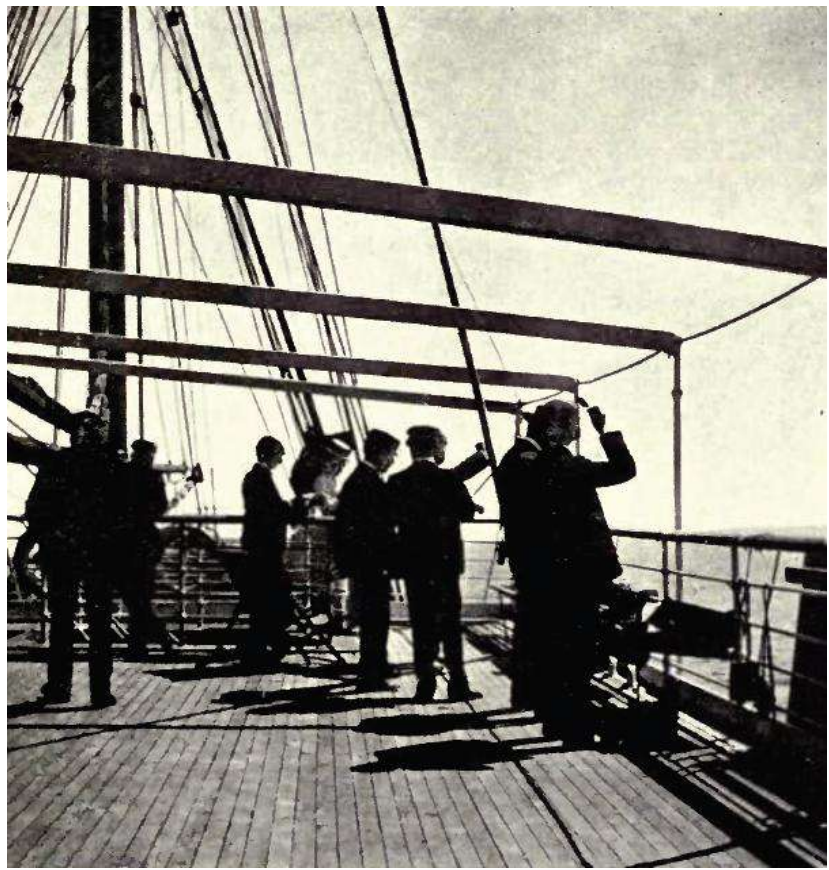

Figure 19: Passengers on the promenade deck of the *SS Austral* observing the partial phase of the 1900 eclipse (93)

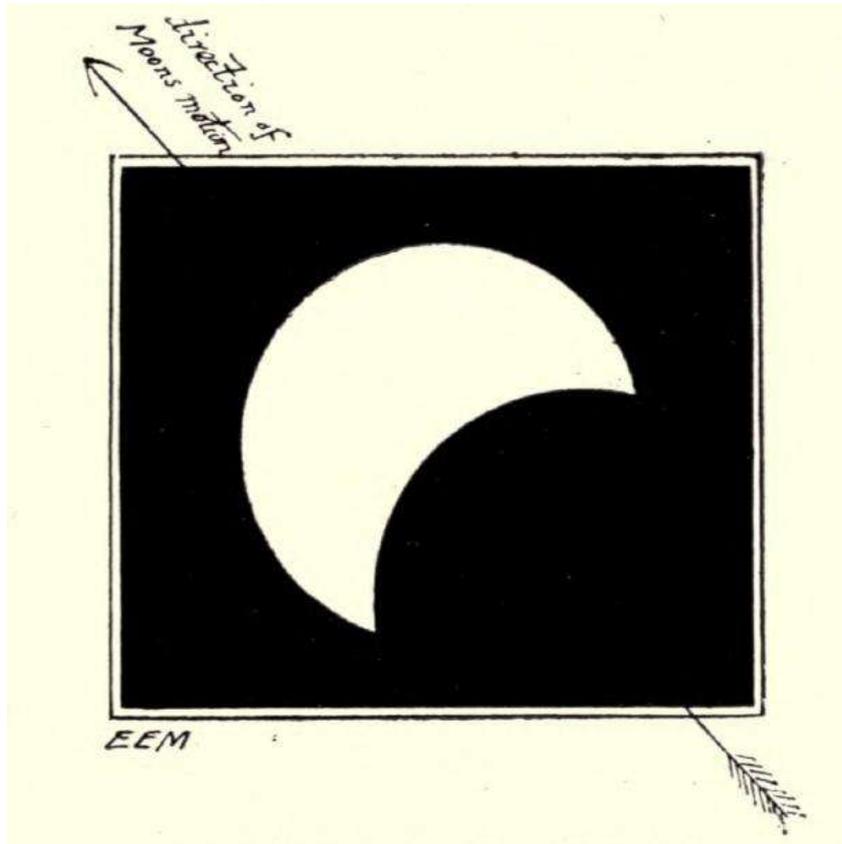

Figure 20: The 1900 solar eclipse: the sun's diameter half obscured
Markwick noted: "By an optical illusion the obscured part looks less than the bright".
The arrow indicates the direction of the Moon's motion
(drawing by E.E. Markwick from ref. (93))

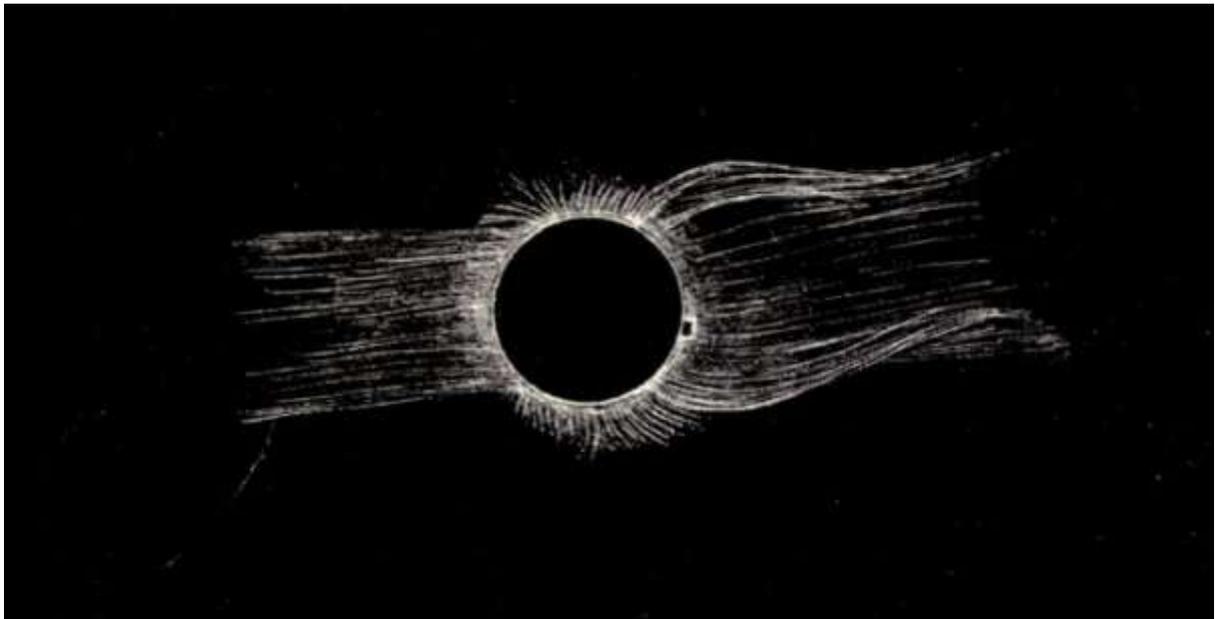

Figure 21: The 1900 solar eclipse: totality
(drawing by E.E. Markwick from ref. (93))

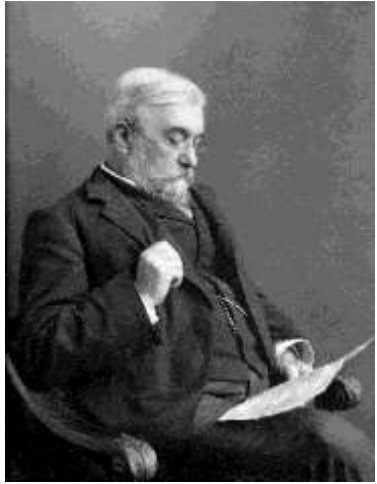

(a)

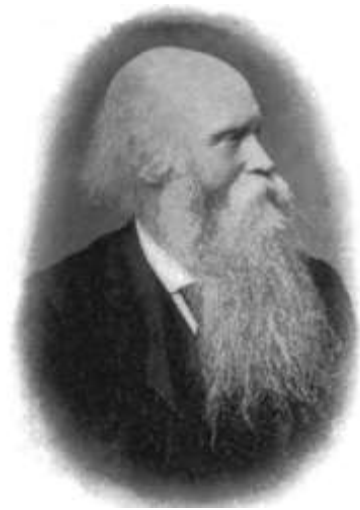

(b)

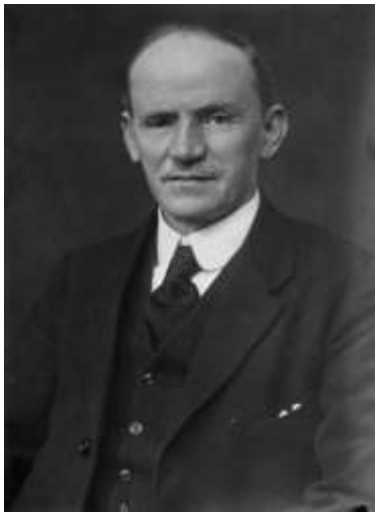

(c)

Figure 22: Astronomers whom Markwick met in Gibraltar after the 1900 solar eclipse (a) Norman Lockyer (1836-1920) (b) Ralph Copeland (1837-1905).Image from ref (138) (c) Alfred Fowler (1868-1940). US Library of Congress

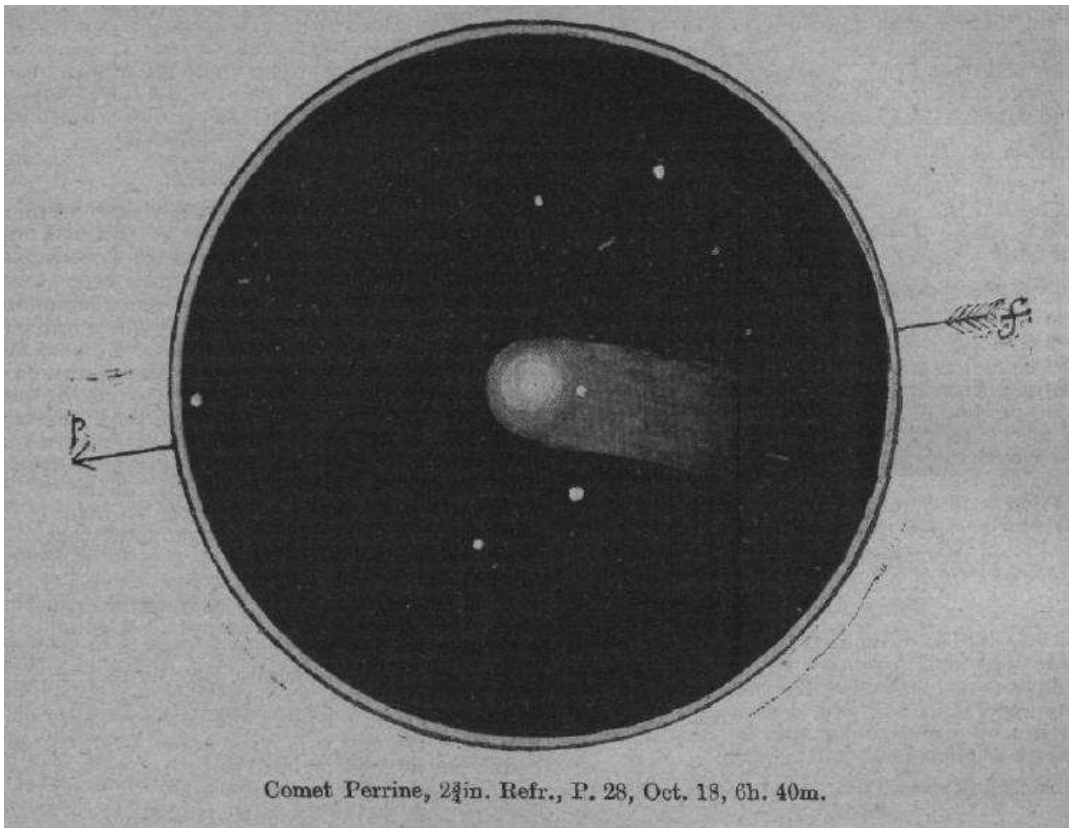

Figure 23: Comet Perrine on 28 Oct 1902 (139)
(E.E. Markwick)

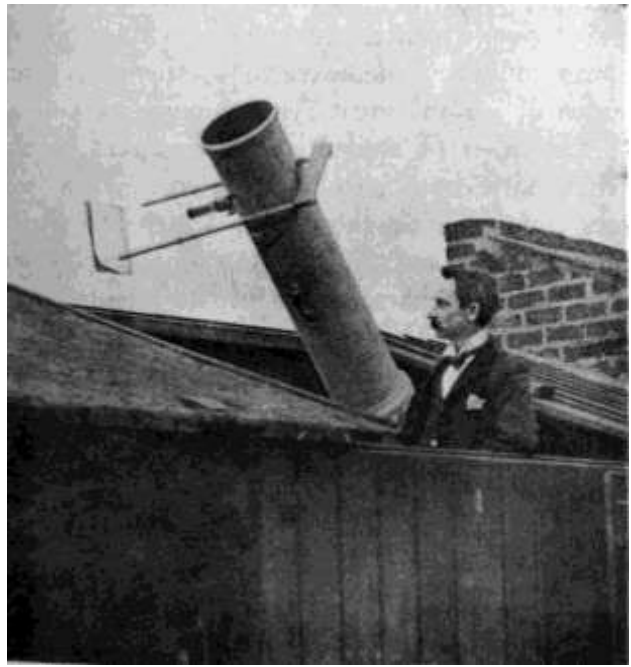

Figure 24: Arthur Mee (1860-1926) using his 8-inch reflector to study sunspots by projection in his Cardiff observatory
(Image courtesy of Bryn Jones)

Figure 25: Advertisement placed by Markwick's wife in the *English Mechanic* for the sale of Markwick's telescopes and observatory (128)

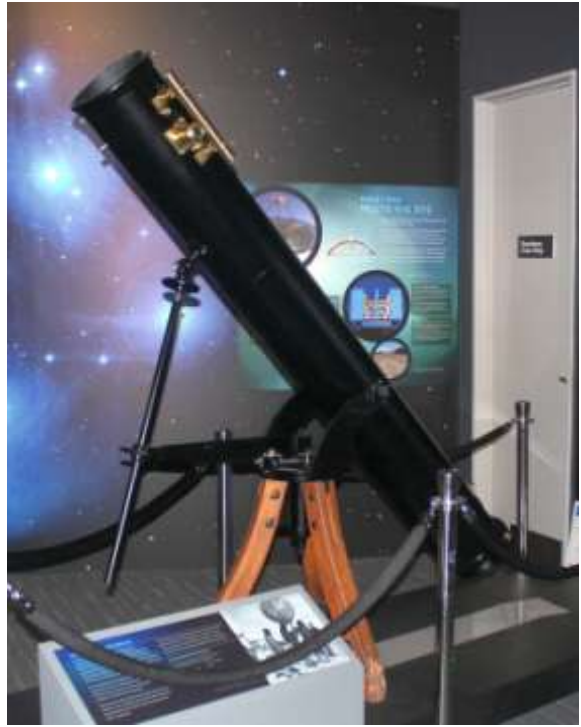

(a)

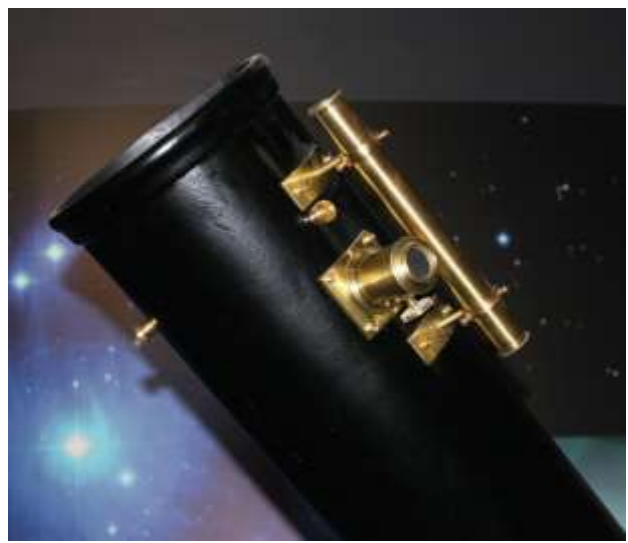

(b)

Figure 26: Markwick's 8½ inch (21.6 cm) Calver reflector
(a) Complete telescope and mounting (b) brass focuser and finder
(Grant Christie, Auckland Observatory)

*Accepted for publication in the Journal of the British Astronomical Association*